\documentclass[12pt]{article}
\usepackage[dvips]{color}
\usepackage{epsfig}
\usepackage{amsmath}
\usepackage{graphicx}

\begin{document}
\title{\bf Interacting Ricci dark energy models with an effective $\Lambda$-term in
Lyra manifold}
\author{{\small {M. Khurshudyan$^{a}$\thanks{Email: khurshudyan@yandex.ru}, \hspace{1mm} J. Sadeghi$^{b}$\thanks{Email: pouriya@ipm.ir},\hspace{1mm} A. Pasqua$^{c}$ \thanks{Email: toto.pasqua@gmail.com},\hspace{1mm} S. Chattopadhyay$^{d}$\thanks{Email: surajcha@iucaa.ernet.in, surajitchatto@outlook.com},\hspace{1mm} R. Myrzakulov$^{e}$\thanks{Email: rmyrzakulov@gmail.com}
\hspace{1mm} and H. Farahani$^{b}$\thanks{Email: h.farahani@umz.ac.ir}}}\\
$^{a}${\small {\em Department of Theoretical Physics, Yerevan State
University,}}\\
{\small {\em 1 Alex Manookian, 0025, Yerevan, Armenia}}\\
$^{b}${\small {\em Department of Physics, Mazandaran University, Babolsar, Iran}}\\
$^{c}${\small {\em Department of Physics, University of Trieste, Via Valerio, 2 34127 Trieste, Italy}}\\
$^{d}${\small {\em Pailan College of Management and Technology, Bengal Pailan Park, Kolkata-700 104, India}}\\
$^{e}${\small {\em Eurasian International Center for Theoretical Physics, Eurasian National University,}}\\
{\small {\em Astana 010008, Kazakhstan}}}  \maketitle
\begin{abstract}
In this paper we consider a universe filled with barotropic dark matter and Ricci dark energy in Lyra's geometry with varying $\Lambda$. We assume two different kinds of interactions between dark matter and dark energy. Then, by using numerical analysis, we investigate some cosmological parameters of the models such as equation of state, Hubble and deceleration parameters.\\\\
\noindent {\bf Keywords:} FRW Cosmology; Ricci Dark Energy; Lyra's Geometry; Varying $\Lambda$.\\\\
{\bf Pacs Number(s):} 95.35.+d, 95.85.-e, 98.80.-k, 95.30.Cq, 97.20.Vs, 98.80.Cq
\end{abstract}

\section{\large{Introduction}}
Dark energy may described accelerated expansion of universe. In that case there are several models for the dark energy such as Einstein's cosmological constant [1] which is the simplest model.  There are also another interesting
models to describe the dark energy such as k-essence model [2], tachyonic models [3], quintessence models [4, 5] and Chaplygin gas models [6-11]. In the another interesting model it is suggested that the energy density proportional to Ricci
scalar could describe an accelerating universe [12].\\
On the other hand, there is possibility to modify the cosmological models by using the Lyras geometry [13]. The effective
cosmological term in Lyras geometry recently has been considered by several authors, such as in Refs. [14, 15]. Also, in the recent work we studied quintessence cosmology with an effective $\Lambda$-term in Lyra manifold. Indeed, we
considered different models by choosing variable $\Lambda$. As we know, the Einstein equations of general relativity do not permit any variations in the cosmological constant, because of the fact that the Einstein tensor has
zero divergence and energy conservation law is also zero. Hence, some modifications of Einstein equations are
necessary. Therefore, the study of the varying $\Lambda$ may be done only through modified field equations
and modified conservation laws. Already we construct several cosmological models based on variation of $G$
and $\Lambda$ [16-18].\\
Now, we would like to investigate another models in the Lyras geometry based on Ricci dark energy. We assumed that the universe filled with barotropic dark matter and Ricci dark energy with possibility of interaction between them. Indeed we will consider three different choices of $\Lambda$ and two different choices of the interaction term. Therefore, totally we have 6 different model and will be able to investigate cosmological parameters of these models numerically. In the next section we introduce our models write explicit forms of interaction terms and varying $\Lambda$, then try to solve corresponding field equations numerically to obtain behavior of some cosmological parameters.

\section{\large{The models}}
We consider 6 different toy models for the universe where an effective energy density and pressure of the fluid which governs dynamics of background assumed to be given as,
\begin{equation}\label{eq:rhoeff}
\rho=\rho_{R}+\rho_{b},
\end{equation}
and,
\begin{equation}\label{eq:Peff}
P=\rho_{R}+\rho_{b},
\end{equation}
where $\rho_{b}$ and $P_{b}$ are energy density and pressure of a barotropic dark matter of the universe with $P_{b}=\omega_{b}\rho_{b}$. Also, Ricci dark energy given by the following energy density [19],
\begin{equation}\label{eq:RDE}
\rho_{R}=3\alpha (\dot{H}+2H^{2}+\frac{k}{a^{2}}),
\end{equation}
where $H$ is the Hubble parameter, and $\alpha$ is a positive constant (wa assume a flat space-time with $k=0$, and $8\pi G = c = 1$). Within an interaction $Q$ between the dark energy and the dark matter we assume a connection between two major components of the Universe. However such assumptions are of a phenomenological origin, which gives a huge number of speculations. Despite to the evidence of an interaction between the dark energy and the dark matter the final form (or forms) of $Q$ is not fixed yet, which is a hard conceptual and theoretical question. Based on the general obvious facts like unit analysis several forms for $Q$ were assumed and considered in literature very intensively. One of the examples could be mentioned an interaction modeled as $Q=3Hb(\rho_{b}+\rho_{R})$, where $b$ is a positive constant. In this work we will consider two different forms of $Q$ as the follows,
\begin{equation}\label{eq:Q1_1}
Q=3Hb(\rho_{R}+\rho_{b})+\gamma \dot{\rho}_{b},
\end{equation}
and,
\begin{equation}\label{eq:Q1_2}
Q=bH^{1-2m}\rho_{b}^{m}\dot{\beta}(t)^{2}.
\end{equation}
We will call the equation (4) as the model 1 and the equation (5) as the model 2. The last form for the interaction (model 2) is motivated by the work \cite{Saridakis}, where $\dot{\phi}^{2}$ is replaces via $\dot{\beta}(t)^{2}$.\\
Also we consider three different models of $\Lambda$. The first and simplest one in constant $\Lambda$ while in the second and third models we assume the following varying $\Lambda$,
\begin{equation}\label{eq:lambda1}
\Lambda=\rho=\rho_{R}+\rho_{b},
\end{equation}
and,
\begin{equation}\label{eq:lambda2_1}
\Lambda=t^{-2}-H^{2}+(\rho_{R}+\rho_{b})e^{-tH}.
\end{equation}
In the next section we recall field equations and then try to solve them numerically.
\section{\large{The field equations}}
Field equations that govern our model of consideration are,
\begin{equation}\label{eq:Einstein eq}
R_{\mu\nu}-\frac{1}{2}g_{\mu\nu}R-\Lambda g_{\mu \nu}+\frac{3}{2}\phi_{\mu}\phi_{\nu}-\frac{3}{4}g_{\mu \nu}\phi^{\alpha}\phi_{\alpha}=T_{\mu\nu}.
\end{equation}
Considering the content of the universe to be a perfect fluid, we have,
\begin{equation}\label{eq:T}
T_{\mu\nu}=(\rho+P)u_{\mu}u_{\nu}-Pg_{\mu \nu},
\end{equation}
where $u_{\mu}=(1,0,0,0)$ is a 4-velocity of the co-moving
observer, satisfying $u_{\mu}u^{\mu}=1$. Let $\phi_{\mu}$ be a time-like
vector field of displacement,
\begin{equation}
\phi_{\mu}=\left ( \frac{2}{\sqrt{3}}\beta,0,0,0 \right ),
\end{equation}
where $\beta=\beta(t)$ is a function of time alone, and the factor $\frac{2}{\sqrt{3}}$ is substituted in order to simplify the writing of all the following equations.
By using FRW metric for a flat Universe,
\begin{equation}\label{s2}
ds^2=-dt^2+a(t)^2\left(dr^{2}+r^{2}d\Omega^{2}\right),
\end{equation}
field equations can be reduced to the following Friedmann equations,
\begin{equation}\label{eq:f1}
3H^{2}-\beta^{2}=\rho+\Lambda,
\end{equation}
and,
\begin{equation}\label{eq:Freidmann2}
2\dot{H}+3H^{2}+\beta^{2}=-P+\Lambda,
\end{equation}
where $H=\frac{\dot{a}}{a}$ is the Hubble parameter, and dot
stands for differentiation with respect to cosmic
time $t$, $d\Omega^{2}=d\theta^{2}+\sin^{2}\theta d\phi^{2}$, also $a(t)$
represents the scale factor. The $\theta$ and $\phi$ parameters are
the usual azimuthal and polar angles of spherical coordinates, with
$0\leq\theta\leq\pi$ and $0\leq\phi<2\pi$. The coordinates ($t, r,
\theta, \phi$) are called co-moving coordinates.\\
The continuity equation reads as,
\begin{equation}\label{eq:coneq}
\dot{\rho}+\dot{\Lambda}+2\beta\dot{\beta}+3H(\rho+P+2\beta^{2})=0.
\end{equation}
Assuming,
\begin{equation}\label{eq:DEDM}
\dot{\rho}+3H(\rho+P)=0,
\end{equation}
then the equation (\ref{eq:coneq}) will give a link between $\Lambda$ and $\beta$ of the following form,
\begin{equation}\label{eq:lbeta}
\dot{\Lambda}+2\beta\dot{\beta}+6H\beta^{2}=0.
\end{equation}
To introduce an interaction between the dark energy and dark matter, Eq. (\ref{eq:DEDM}) we should mathematically split it into two following equations,
\begin{equation}\label{eq:inteqm}
\dot{\rho}_{DM}+3H(\rho_{DM}+P_{DM})=Q,
\end{equation}
and,
\begin{equation}\label{eq:inteqG}
\dot{\rho}_{DE}+3H(\rho_{DE}+P_{DE})=-Q.
\end{equation}
For the barotropic fluid with $P_{b}=\omega_{b}\rho_{b}$ and using the equation (\ref{eq:inteqm}), the dynamics of energy density for the interaction term (4) will take the following form,
\begin{equation}\label{19}
(1-\gamma)\dot{\rho}_{b}+3H(1+\omega_{b}-b)\rho_{b}=3Hb\rho_{R},
\end{equation}
while for the interaction term (5) we will have,
\begin{equation}\label{20}
\dot{\rho}_{b}+3H(1+\omega_{b}-bH^{2(1-m)}\rho_{b}^{m-1}\dot{\beta}^{2})\rho_{b}=0,
\end{equation}
where $\rho_{b}$ stands for dark matter density and $\rho_{R}$ stands for Ricci dark energy density. The equation (\ref{eq:inteqG}) will allow us to find pressure of Ricci dark energy. Cosmological parameters of our interest are EoS parameters of each fluid components $\omega_{i}=P_{i}/\rho_{i}$, EoS parameter of composed fluid,
\begin{equation}\label{21}
\omega_{tot}=\frac{P_{b}+P_{R} }{\rho_{b}+\rho_{R}},
\end{equation}
and deceleration parameter $q$, which can be written as,
\begin{equation}\label{eq:accchange}
q=\frac{1}{2}(1+3\frac{P}{\rho} ).
\end{equation}
\section{\large{The case of constant $\Lambda$}}
We will start the analyze from the model with the constant $\Lambda$.
According to this assumption, Eq. (\ref{eq:coneq}) will be modified as follow,
\begin{equation}\label{23}
\dot{\rho}+2\beta\dot{\beta}+3H(\rho+P+2\beta^{2})=0.
\end{equation}
Using the equation (16) we have,
\begin{equation}
\dot{\beta}+3H\beta=0.
\end{equation}
Integration of the last equation gives relation between $\beta(t)$ and $a(t)$ as follow,
\begin{equation}
\beta=\beta_{0}a^{-3}.
\end{equation}
Numerical analysis of the model summarized in the following subsections corresponding to shape of the interaction term. For both models with constant $\Lambda$ we took $\Lambda=0.0, 0.5, 0.7, 1.0, 1.2$ corresponding to from purple to black lines on the plots.
\subsection{\large{The model 1}}
In the first model we use the interaction term given by the equation (4).\\
Plots of Fig. 1 show time evolution of Hubble expansion parameter (top panels) and the deceleration parameter (bottom panels) for the selected values of $\alpha$, $b$ and $\gamma$. First of all we can see that the case of $\Lambda=0$ for small value of $\alpha$ (such as $\alpha=0.3$), Hubble parameter is increasing function of time which is unexpected. Therefore the value of $\alpha$ should be larger such as $\alpha=0.75$ (for example). Then we can see that increasing interaction term decreases value of $H$. However we expect that it's value decreased to a constant, which satisfied for example by choosing $\alpha=0.5$, $b=0.1$ and $\gamma=0.2$.\\
On the other hand other plots show that the deceleration parameter takes negative value at the early universe with $q<-1$ and yield to $q\geq-1$ at the late time. If we seek a model to reach $q\rightarrow-1$, then we can choose $\alpha=0.5$, $b=0.1$ and $\gamma=0.2$ as previous.\\
Plots of Fig. 2 represent behavior of total EoS (top panels) and Ricci dark energy EoS (bottom panels). We can see that behavior of EoS parameter strongly depend on parameters of the model. However by choosing appropriate values for $\alpha$, $b$ and $\gamma$ we can obtain $\omega\rightarrow-1$ which is expected. For the larger values of $\alpha$ the total EoS parameter is increasing function of time, but choosing smaller values of $\alpha$ shows that the total EoS parameter may be decreasing function of time which may yields to -1 at the late time. The best choice to obtain this may be $\alpha=0.5$, $b=0.1$ and $\gamma=0.2$ as before. The Ricci dark energy EoS parameter also can yields -1 for the mentioned parameters. Generally it behaves as $\omega_{R}\leq-1$ for the small values of parameters, therefore universe is in phantom phase which consistent with the recent observations. Otherwise the large values of $\alpha$, $b$ and $\gamma$ gives $\omega_{R}>-1$ at the late time which is characteristic of quintessence-like universe.

\begin{figure}[h!]
 \begin{center}$
 \begin{array}{cccc}
\includegraphics[width=50 mm]{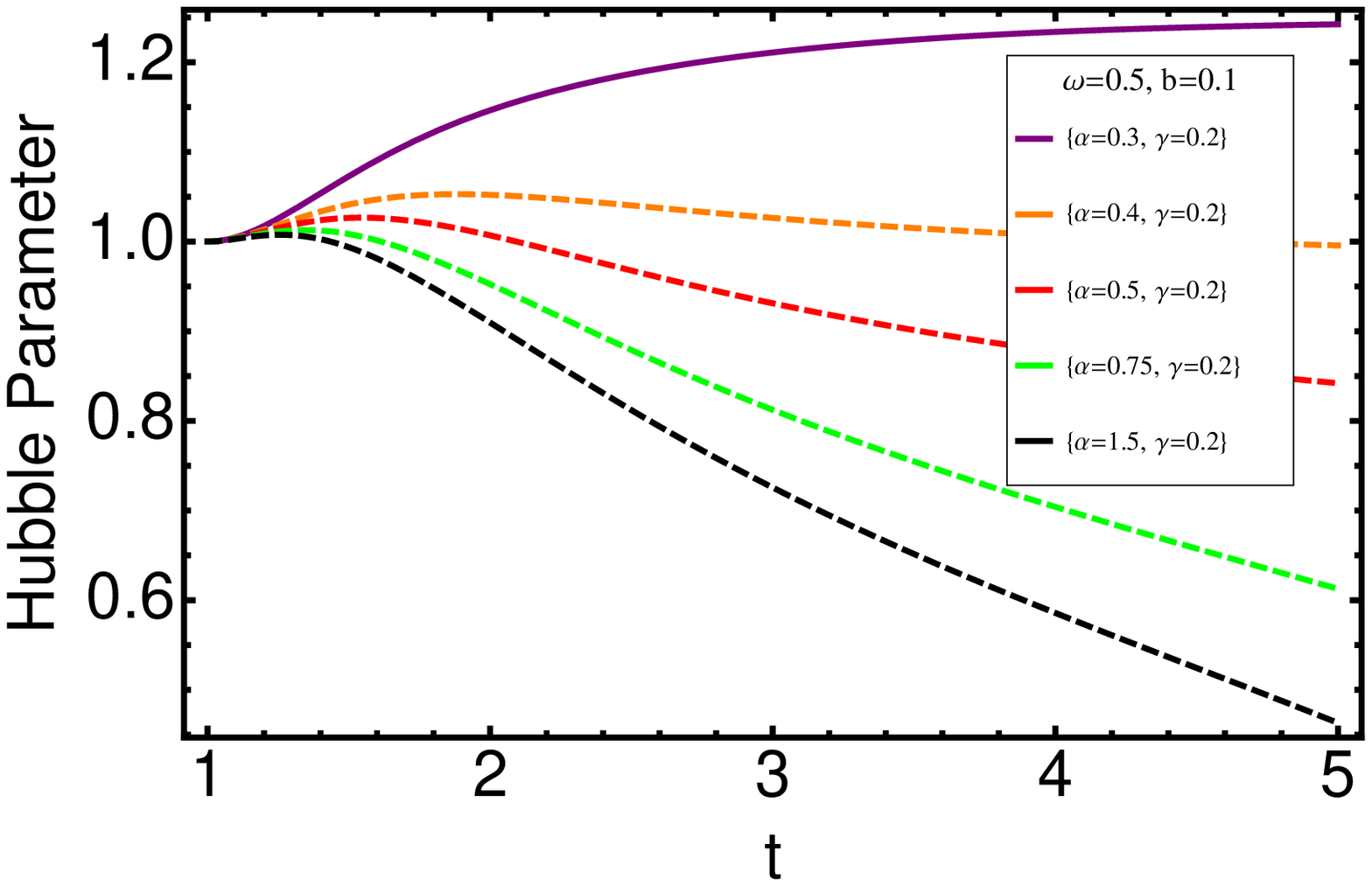} &
\includegraphics[width=50 mm]{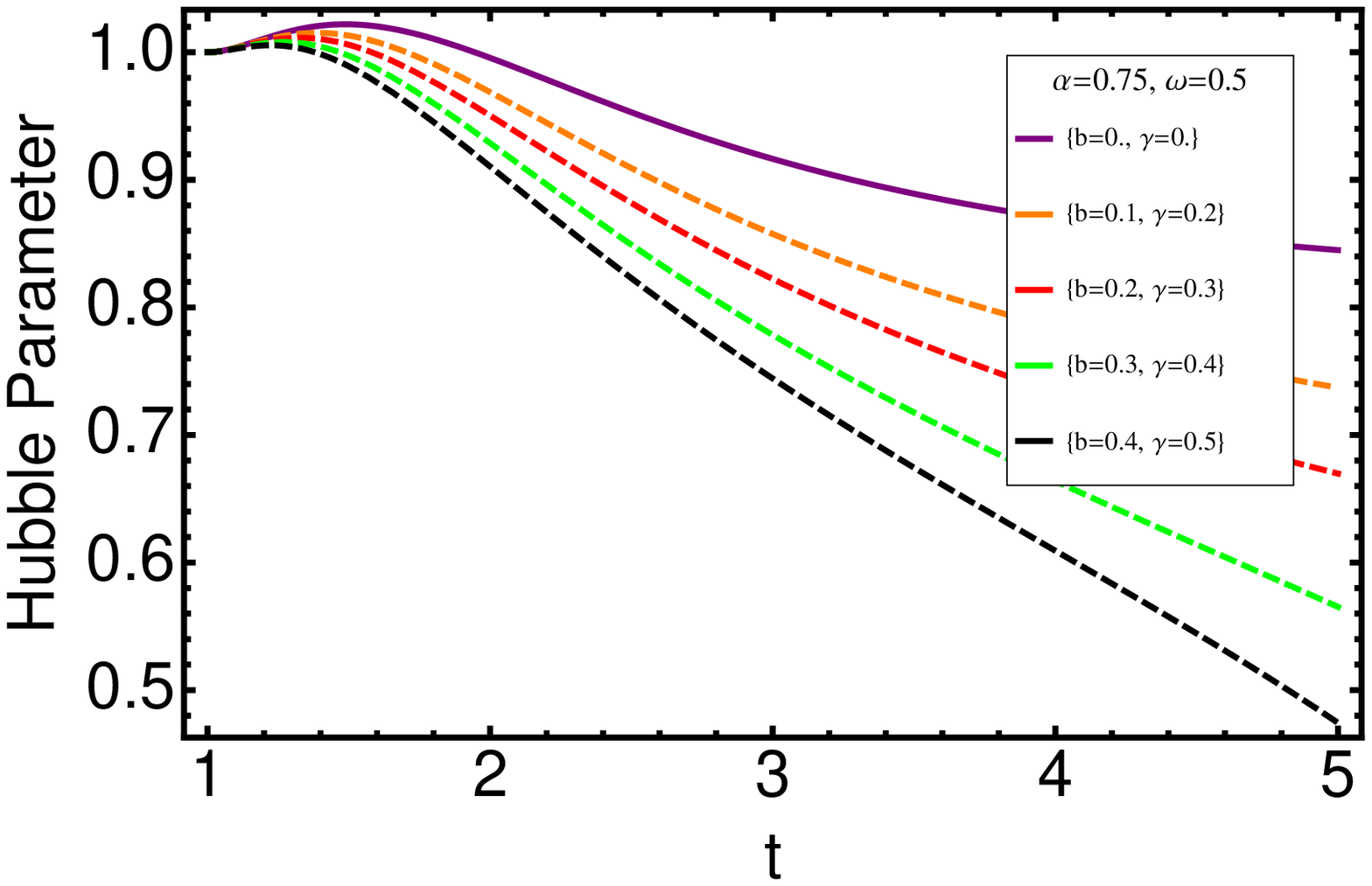} \\
\includegraphics[width=50 mm]{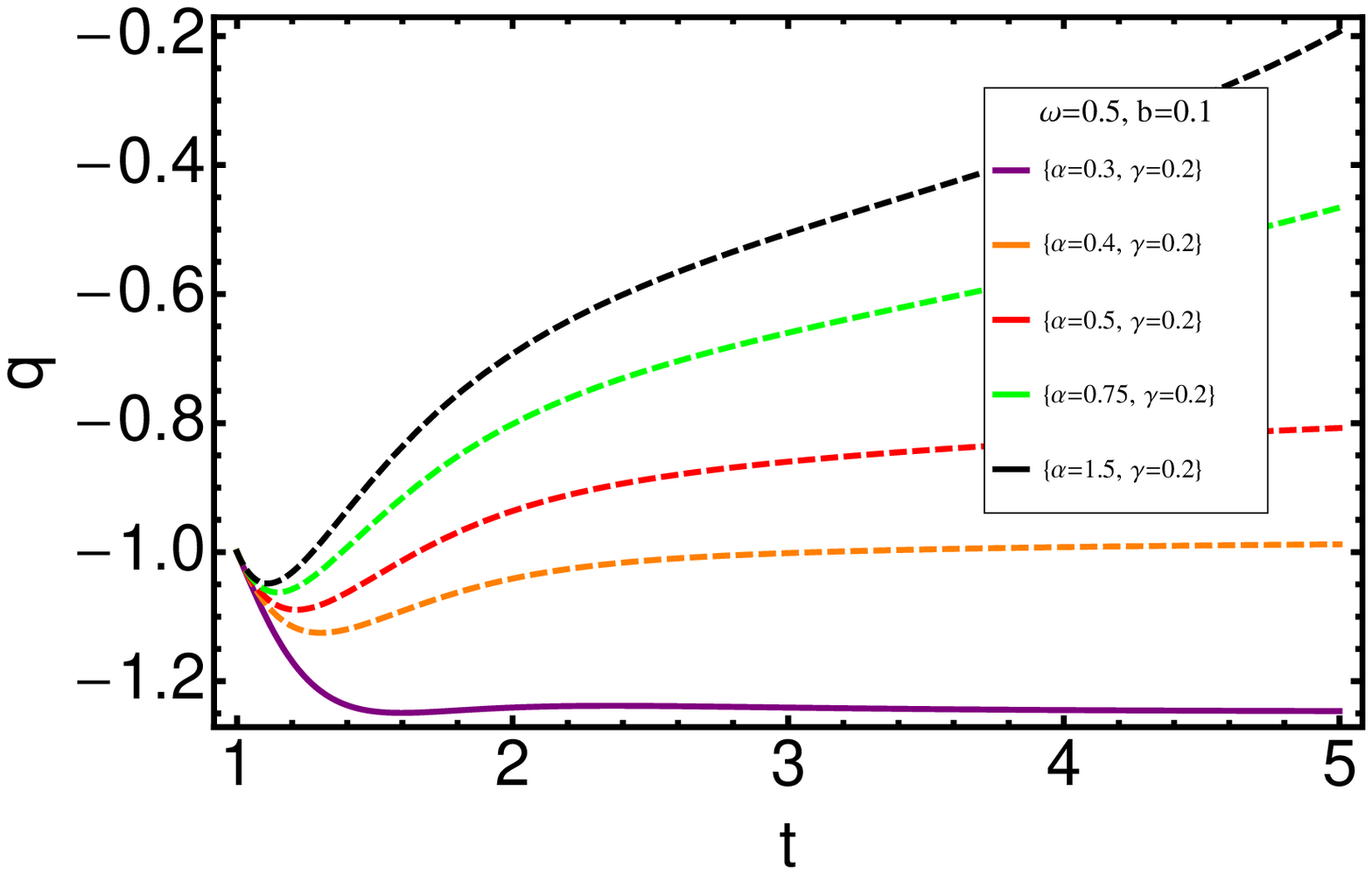} &
\includegraphics[width=50 mm]{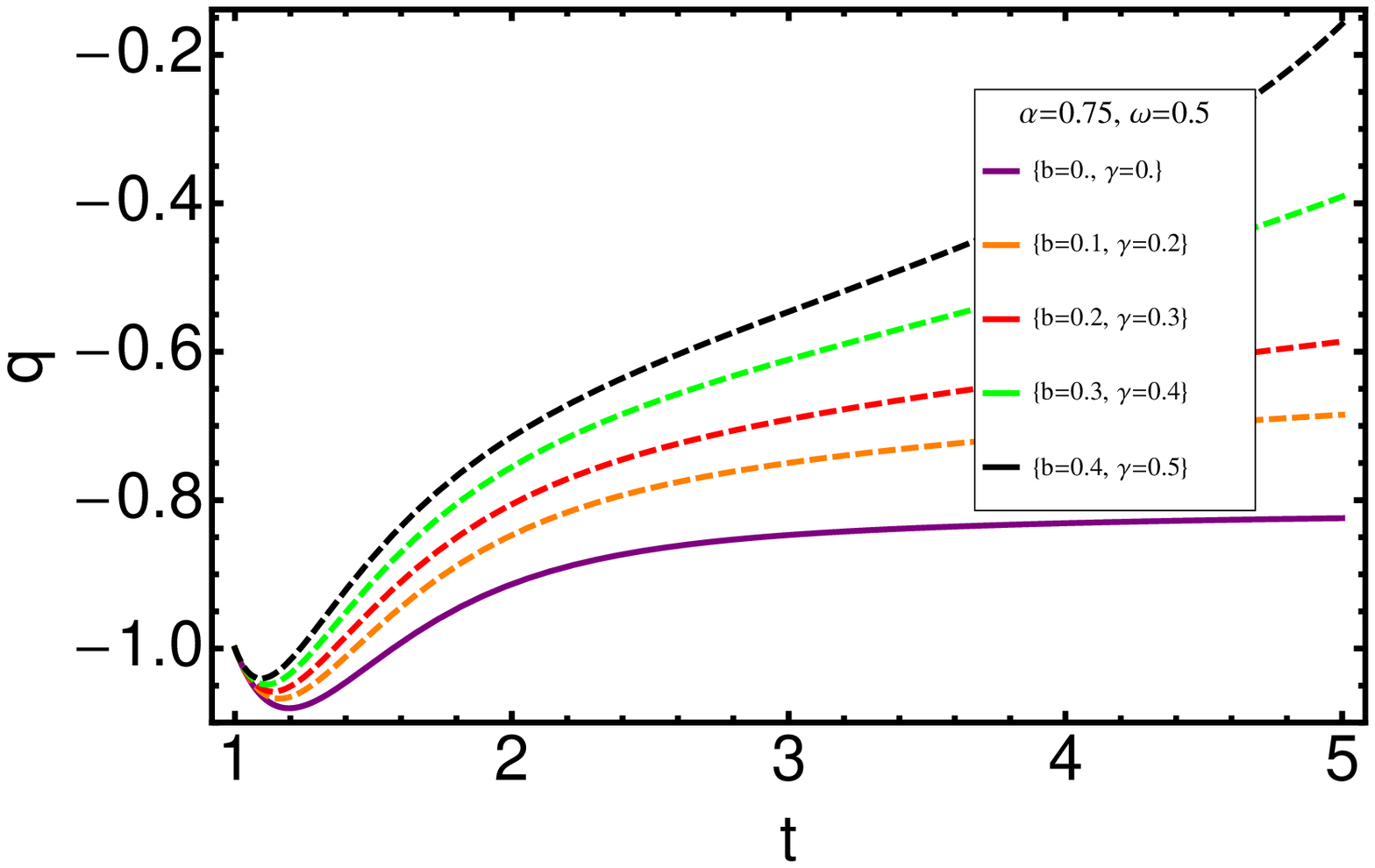}&
 \end{array}$
 \end{center}
\caption{Behavior of Hubble parameter $H$ and $q$ against $t$ for the constant $\Lambda$ and model 1.}
 \label{fig:const1_1}
\end{figure}

\begin{figure}[h!]
 \begin{center}$
 \begin{array}{cccc}
\includegraphics[width=50 mm]{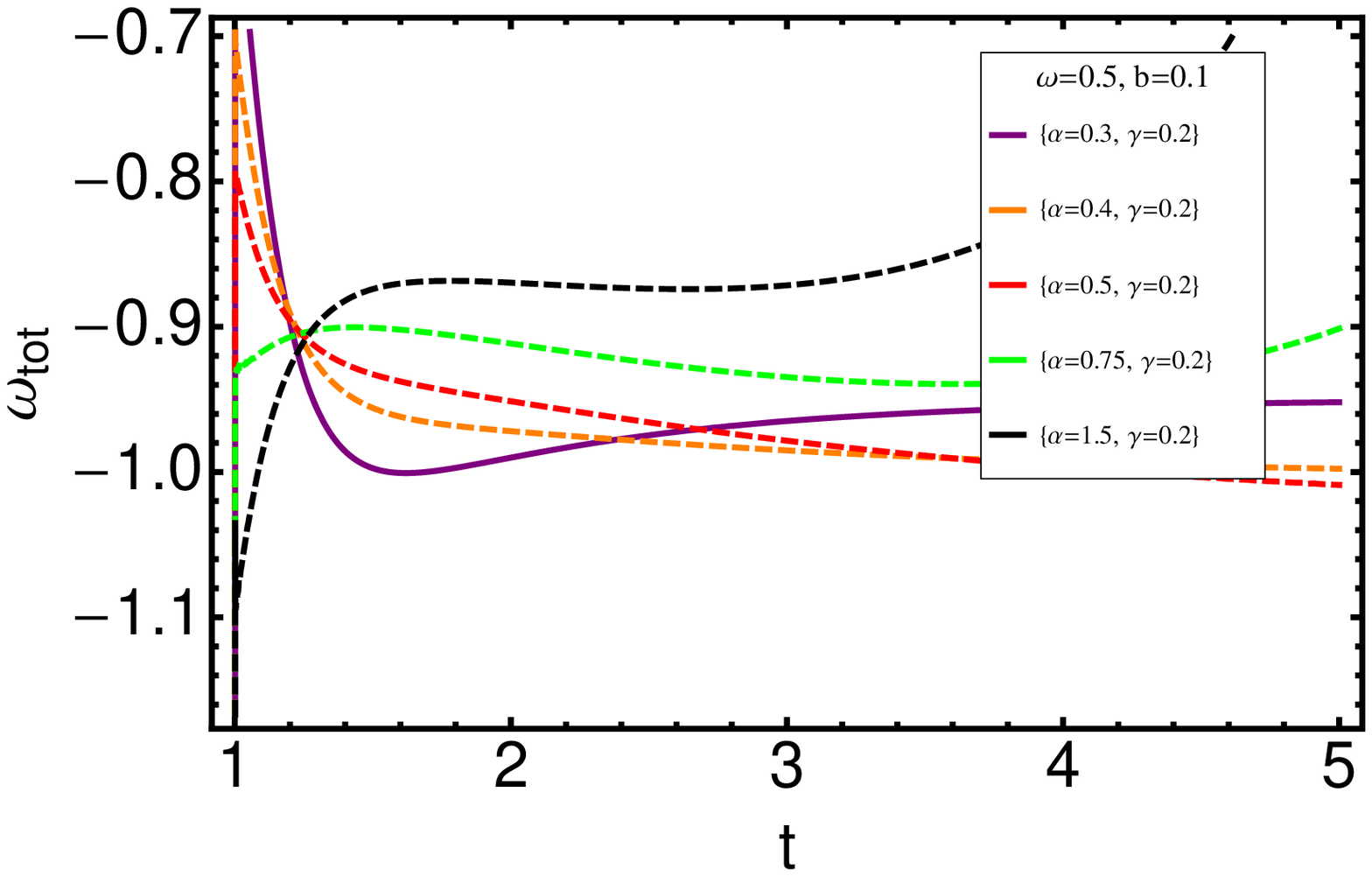} &
\includegraphics[width=50 mm]{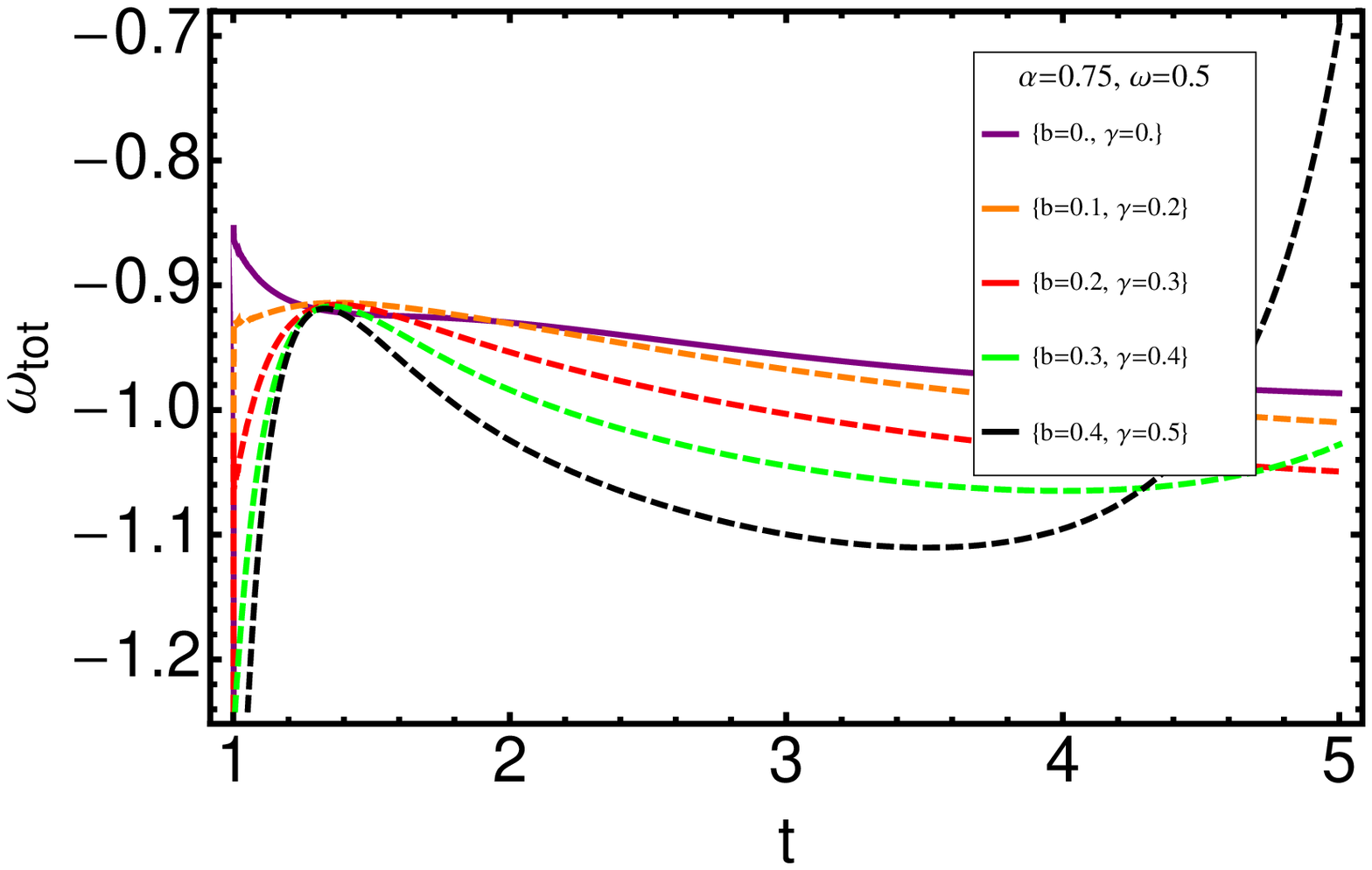}\\
\includegraphics[width=50 mm]{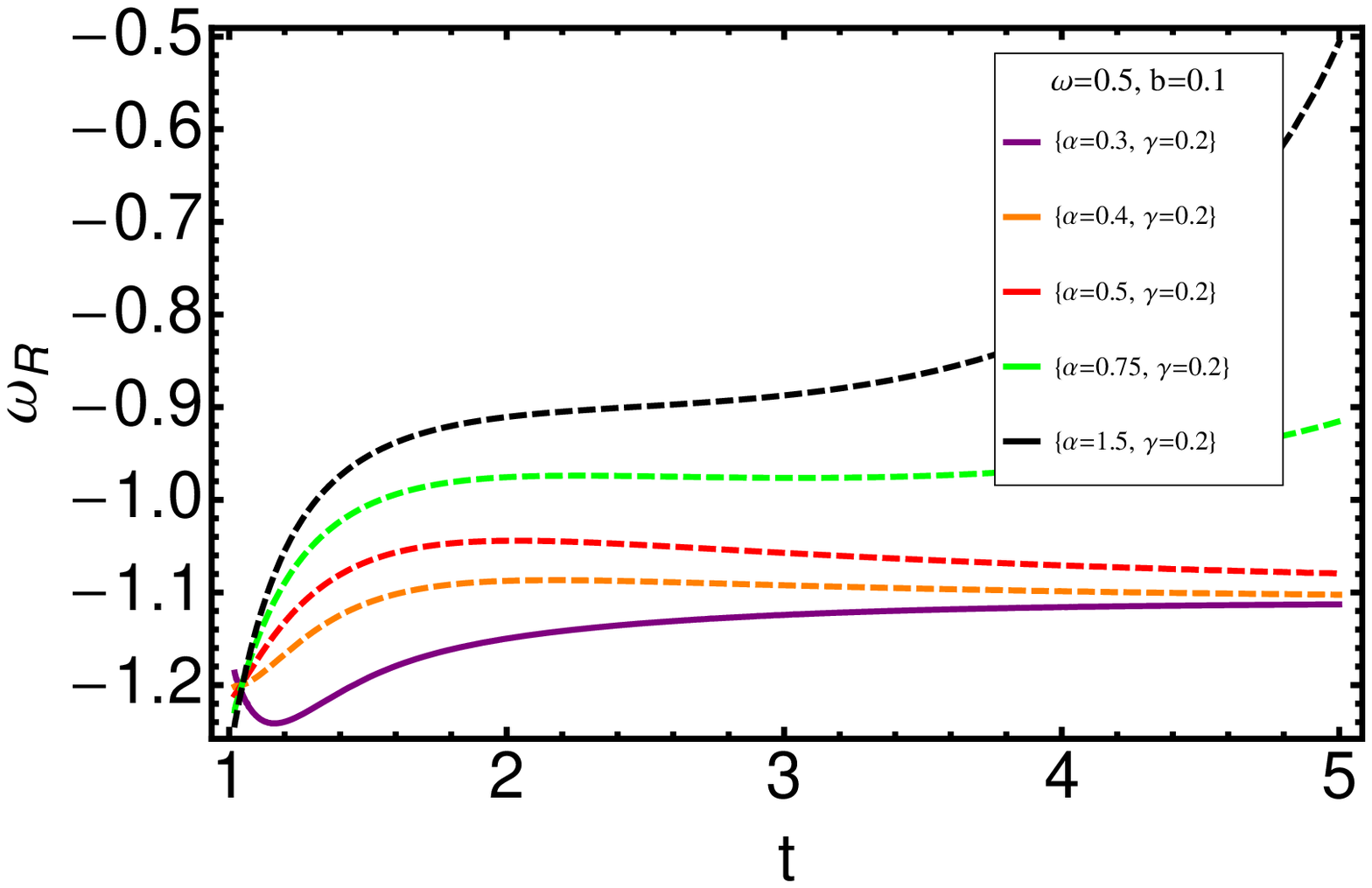} &
\includegraphics[width=50 mm]{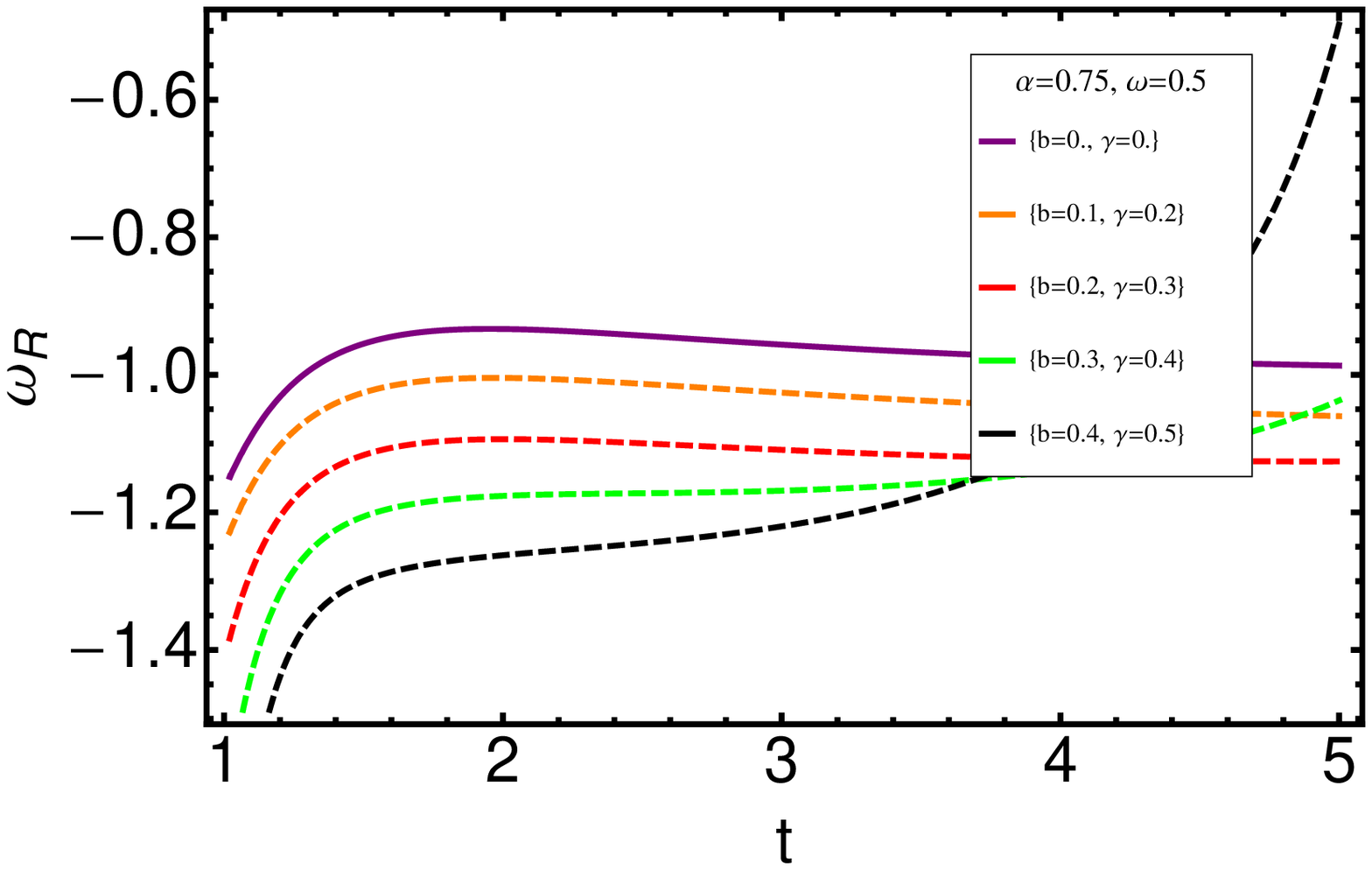}
 \end{array}$
 \end{center}
\caption{Behavior of filed $\omega_{tot}$ and $\omega_{R}$ against $t$ for the constant $\Lambda$ and model 1.}
 \label{fig:const1_2}
\end{figure}

\subsection{\large{The model 2}}
In the second model we use the interaction term given by the equation (5). We vary the parameters $b$ and $m$ to find that the Hubble expansion parameter is totally decreasing function of time for all values of $b$ and $m$ (see top plots of Fig. 3). Also we find that increasing $b$ and $m$ (increasing strength of interaction) decreases value of the Hubble expansion parameter.\\
Other plots of Fig. 3 show that the deceleration parameter decreased with time at the early universe and then grow up to reach -1 for non-interacting case while $q>-1$  for the interacting case. For example, $m=2$, $\alpha=0.75$, and $0.05\leq b\leq0.1$ give us $q\sim -0.6$ at the late time which agree with some observational data.\\

\begin{figure}[h!]
 \begin{center}$
 \begin{array}{cccc}
\includegraphics[width=50 mm]{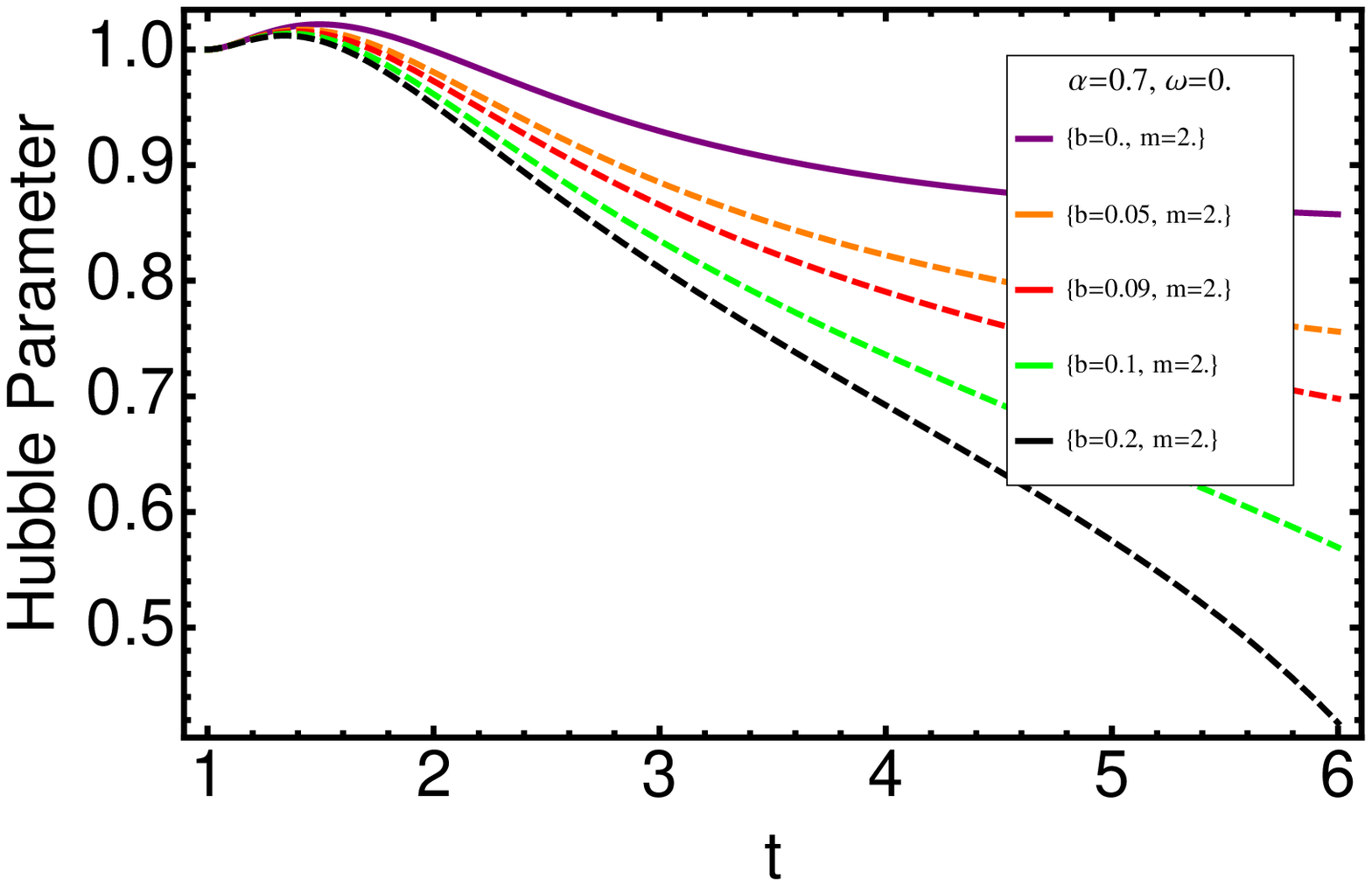}&
\includegraphics[width=50 mm]{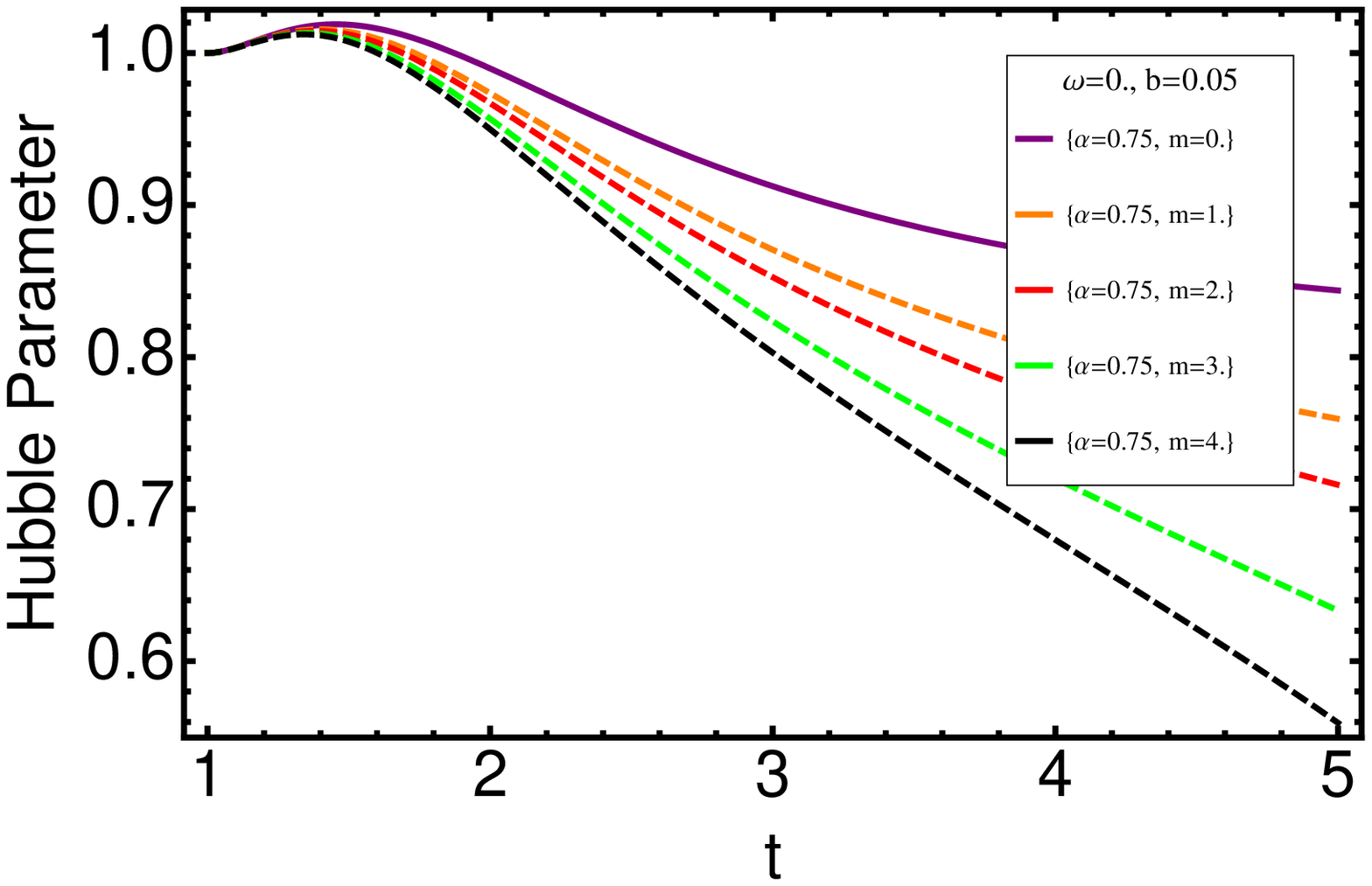} \\
\includegraphics[width=50 mm]{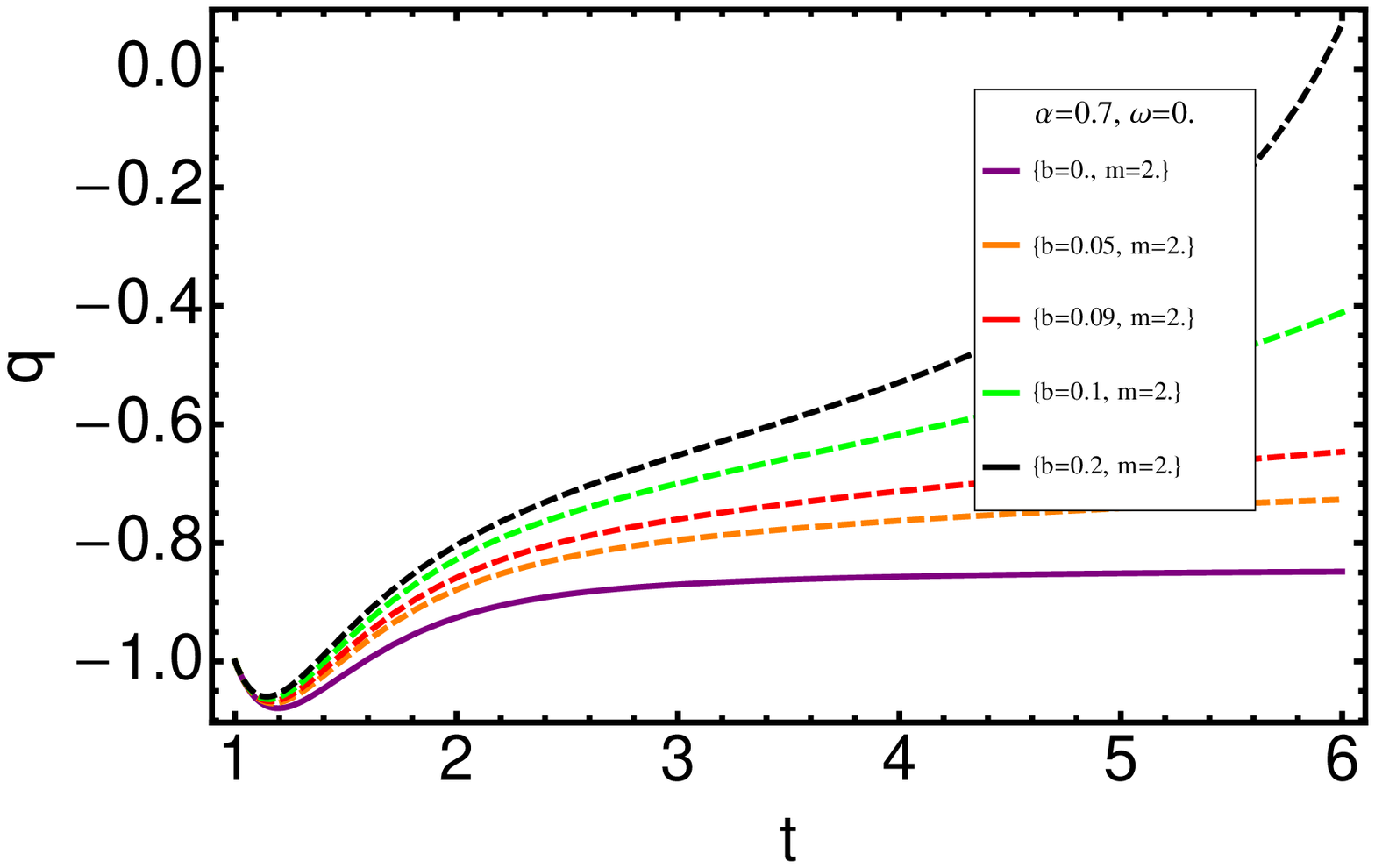}&
\includegraphics[width=50 mm]{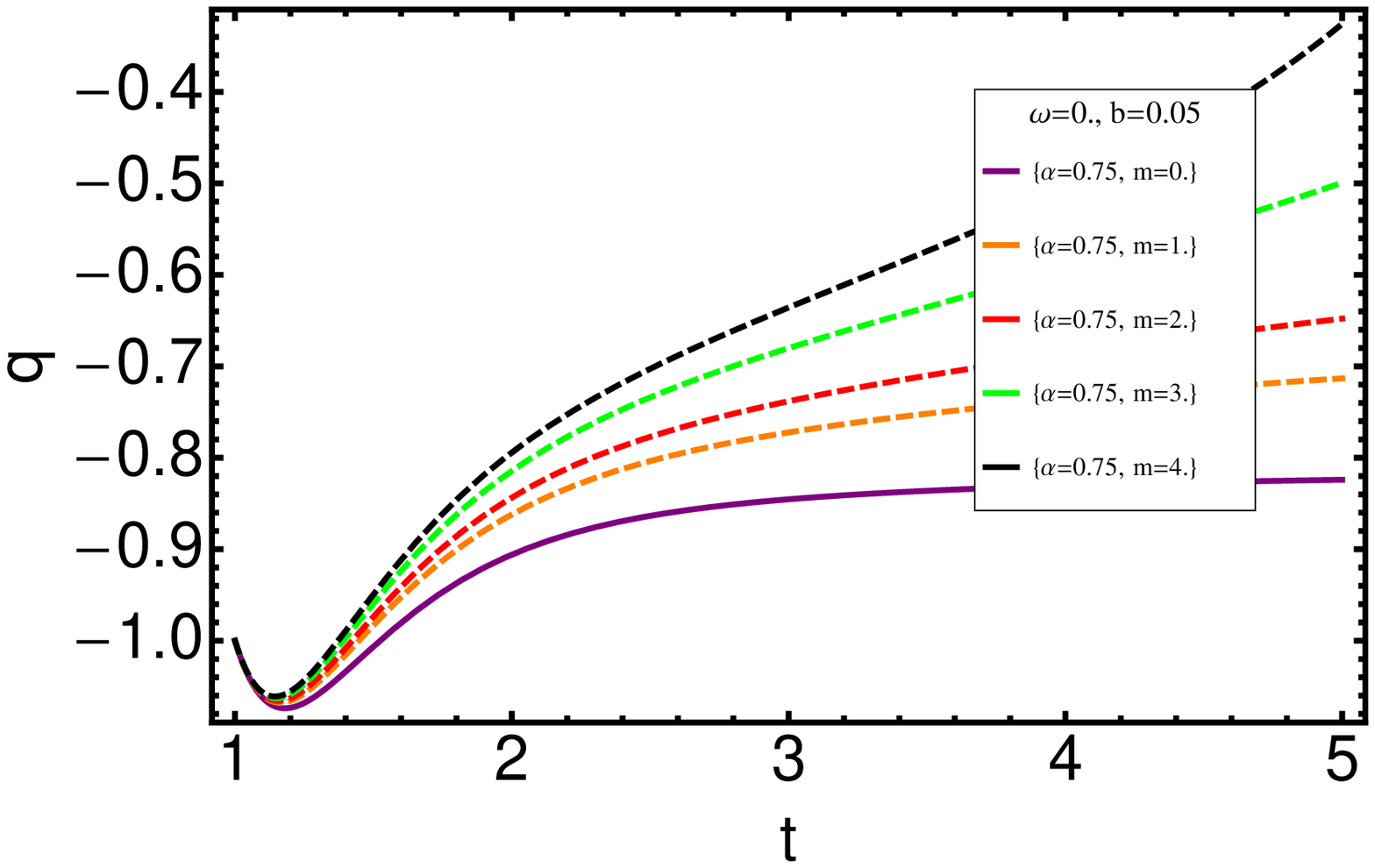}
 \end{array}$
 \end{center}
\caption{Behavior of Hubble parameter $H$ and $q$ against $t$ for the constant $\Lambda$ and model 2.}
 \label{fig:const2_1}
\end{figure}

\begin{figure}[h!]
 \begin{center}$
 \begin{array}{cccc}
\includegraphics[width=50 mm]{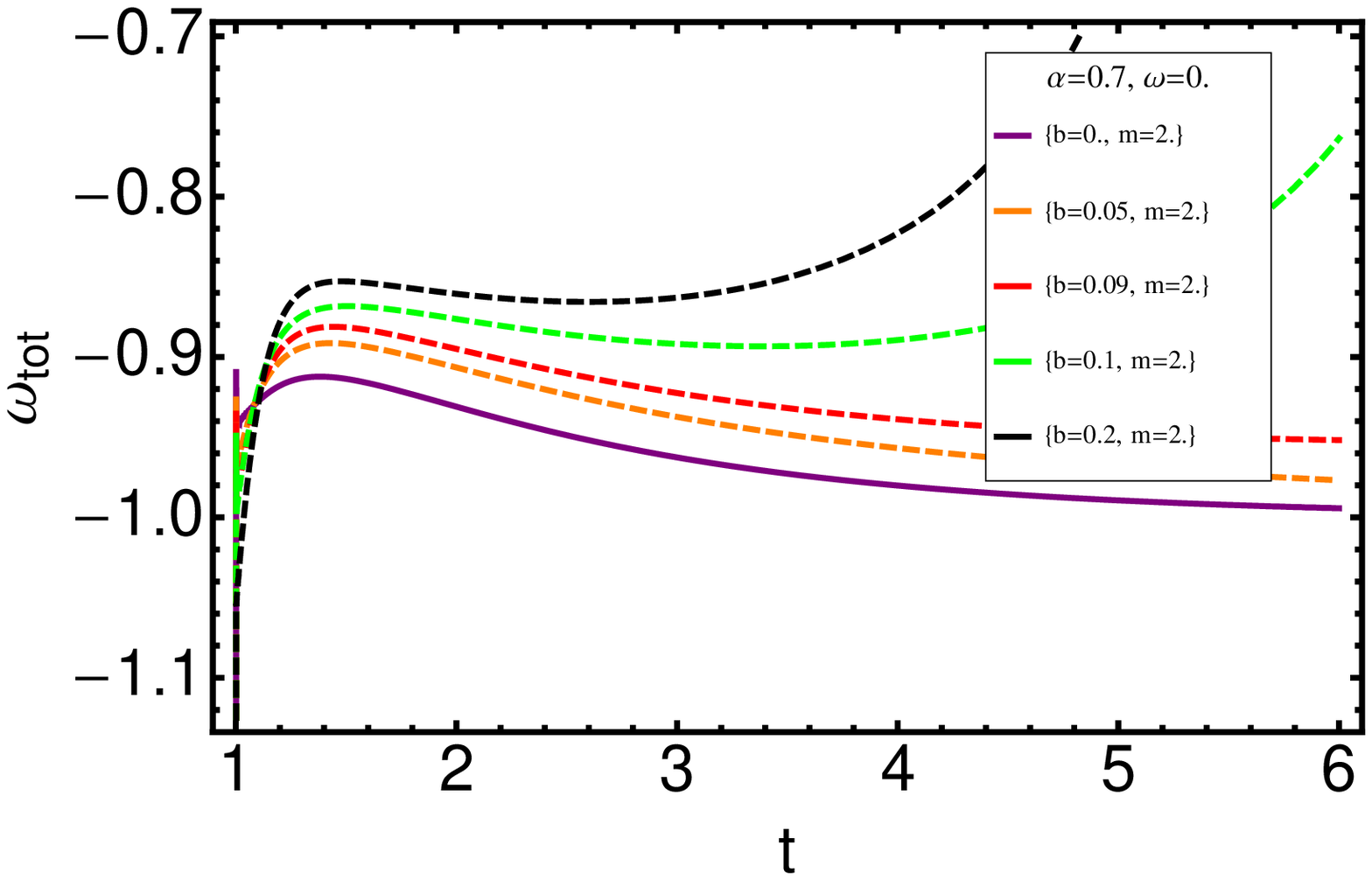}&
\includegraphics[width=50 mm]{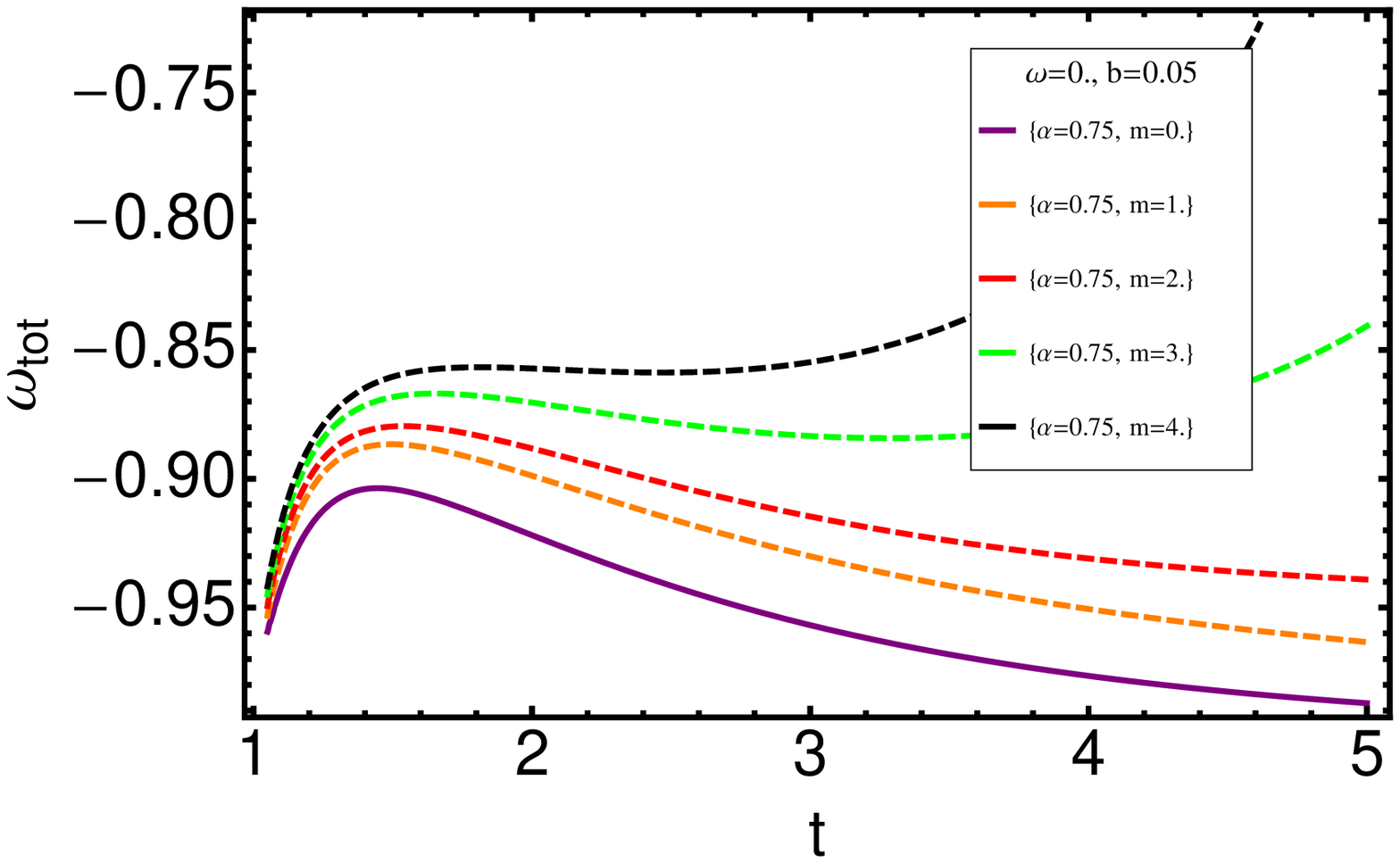}\\
\includegraphics[width=50 mm]{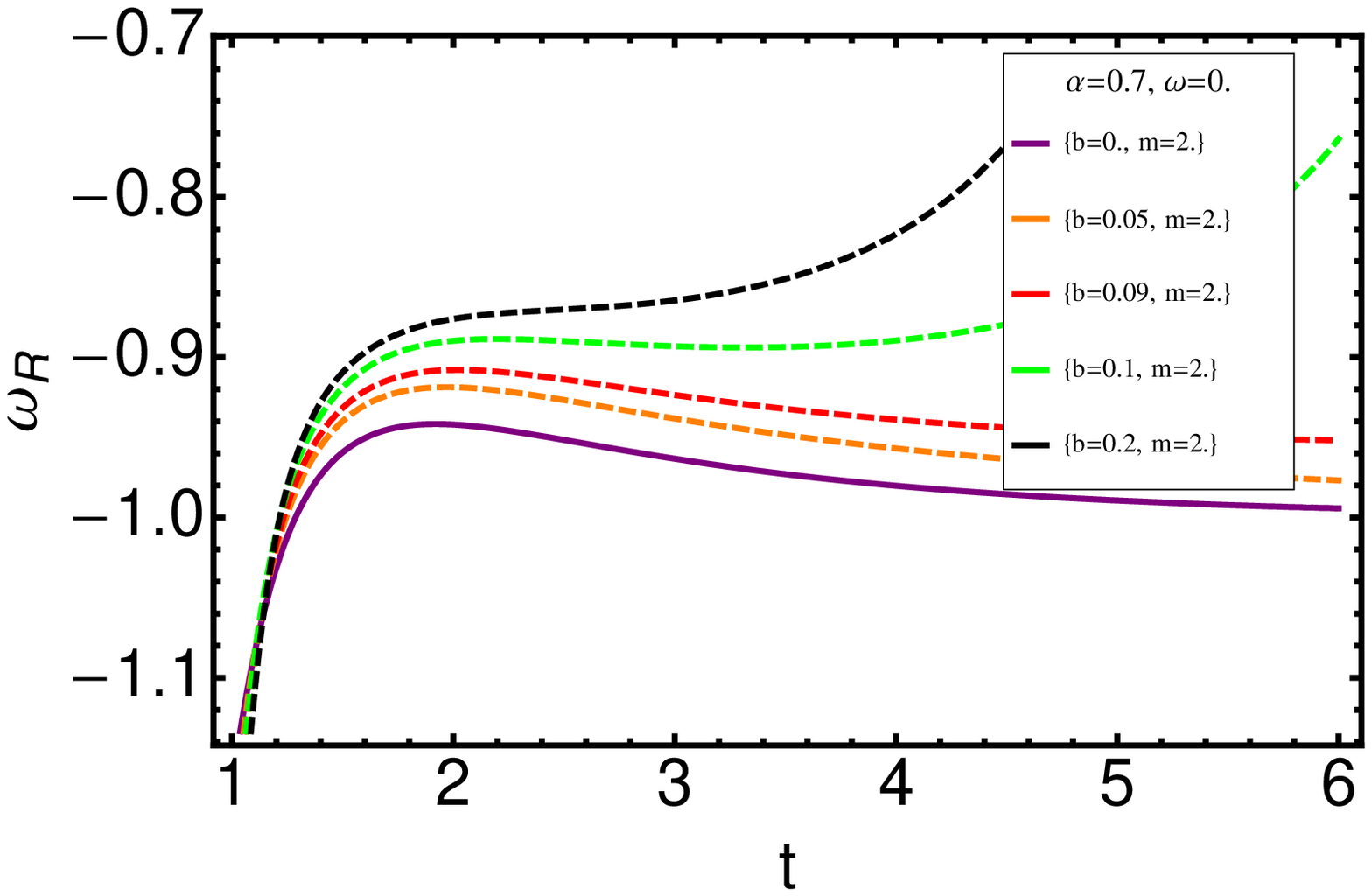}&
\includegraphics[width=50 mm]{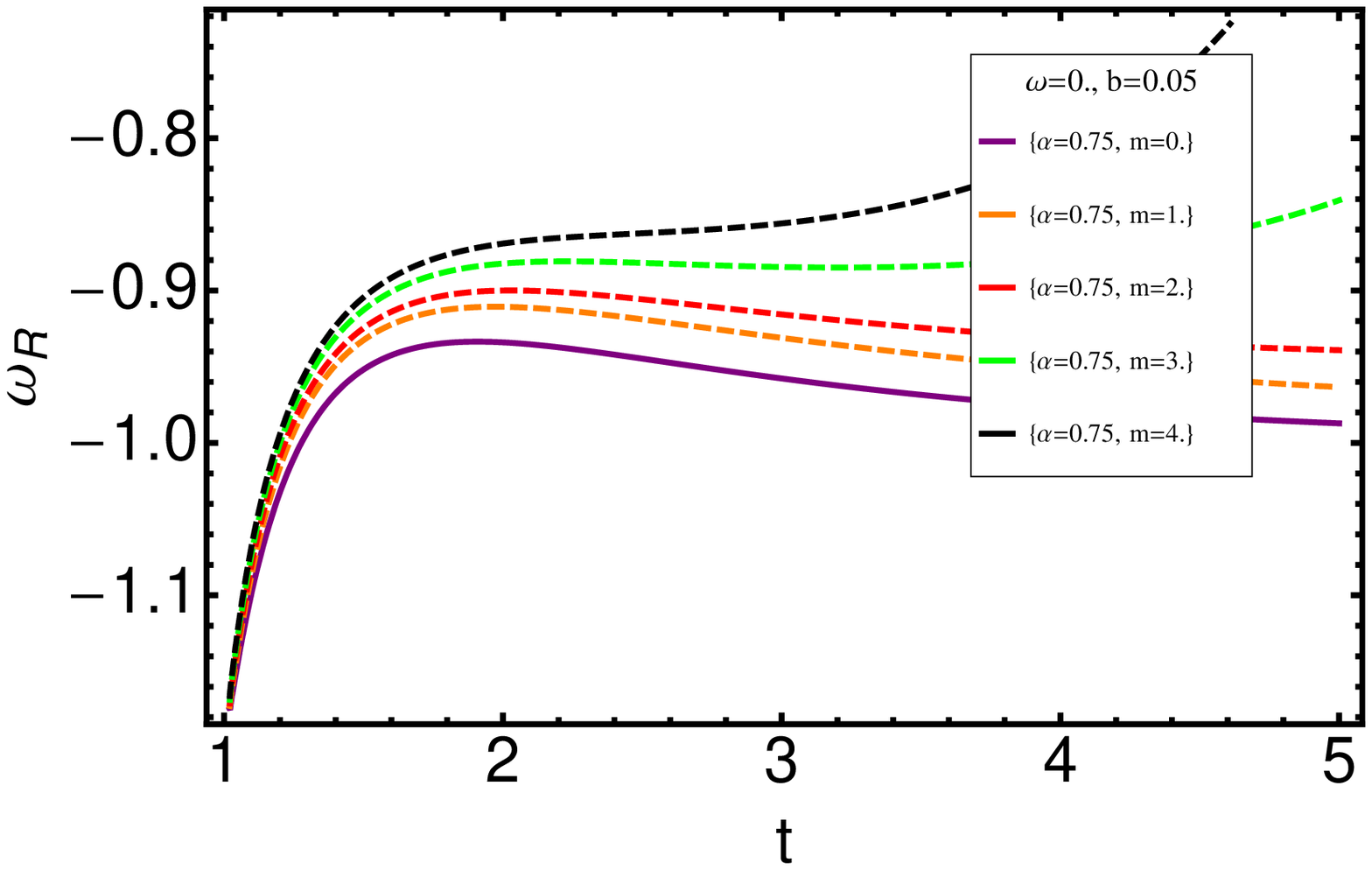}
 \end{array}$
 \end{center}
\caption{Behavior of filed $\omega_{tot}$ and $\omega_{R}$ against $t$ for the constant $\Lambda$ and model 2.}
 \label{fig:const2_2}
\end{figure}

Plots of Fig. 4 show that time evolution of EoS for the cases of non-interaction is faster than the cases with interaction. However for the small values of the parameters $b$ and $m$ (for example $m<3$ and $b<0.09$) the total EoS yields to -1 at the late time. Also Ricci dark energy EoS is smaller than -1 at the early universe but grater than -1 at the other time.

\section{\large{The case of varying $\Lambda$ proportional to the total density}}
In this section we assume $\Lambda$ given by the equation (6) which is proportional to the total density. It may have large value at the early universe which decreases with time to infinitesimal value today.
\subsection{\large{The model 1}}
We obtained behavior of Hubble expansion parameter and the deceleration parameter numerically in Fig. 5. We can see that the Hubble parameter increased with time at the early universe, then decreased to reach a constant at present epoch. It is clear that increasing $\alpha$, $b$ and $\gamma$ decrease value of $H$. We also can see that the deceleration parameter is totally negative which yield to constant value at the late time, which expected. Combining BAO/CMB observations with SNIa data processed with the MLCS2k2 light-curve fitter suggest $q\sim-0.53$ [21] which can be obtained in this model by choosing $\alpha=0.75$, $b=0.1$ and $\gamma=0.2$.\\

\begin{figure}[h!]
 \begin{center}$
 \begin{array}{cccc}
\includegraphics[width=50 mm]{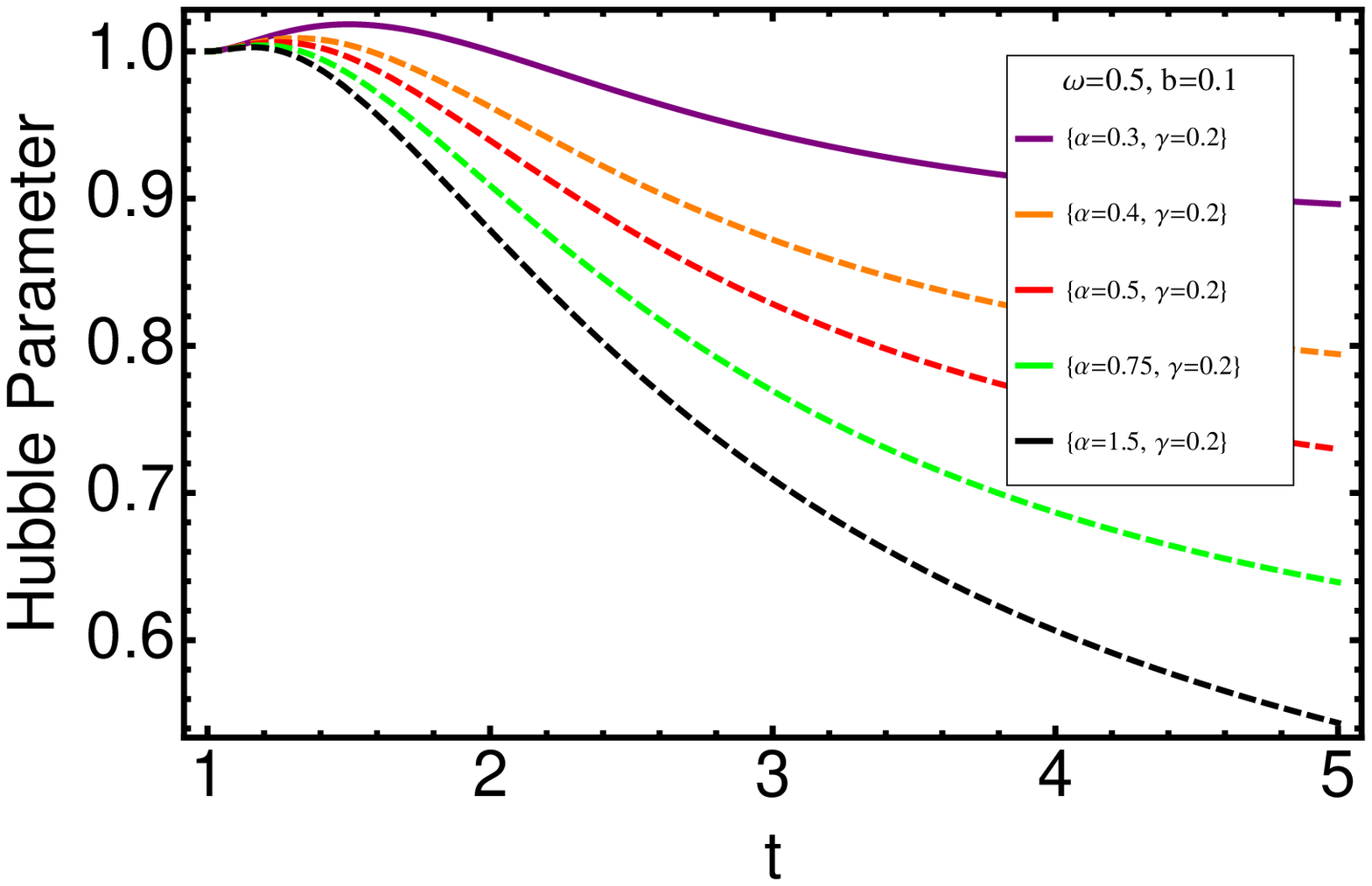} &
\includegraphics[width=50 mm]{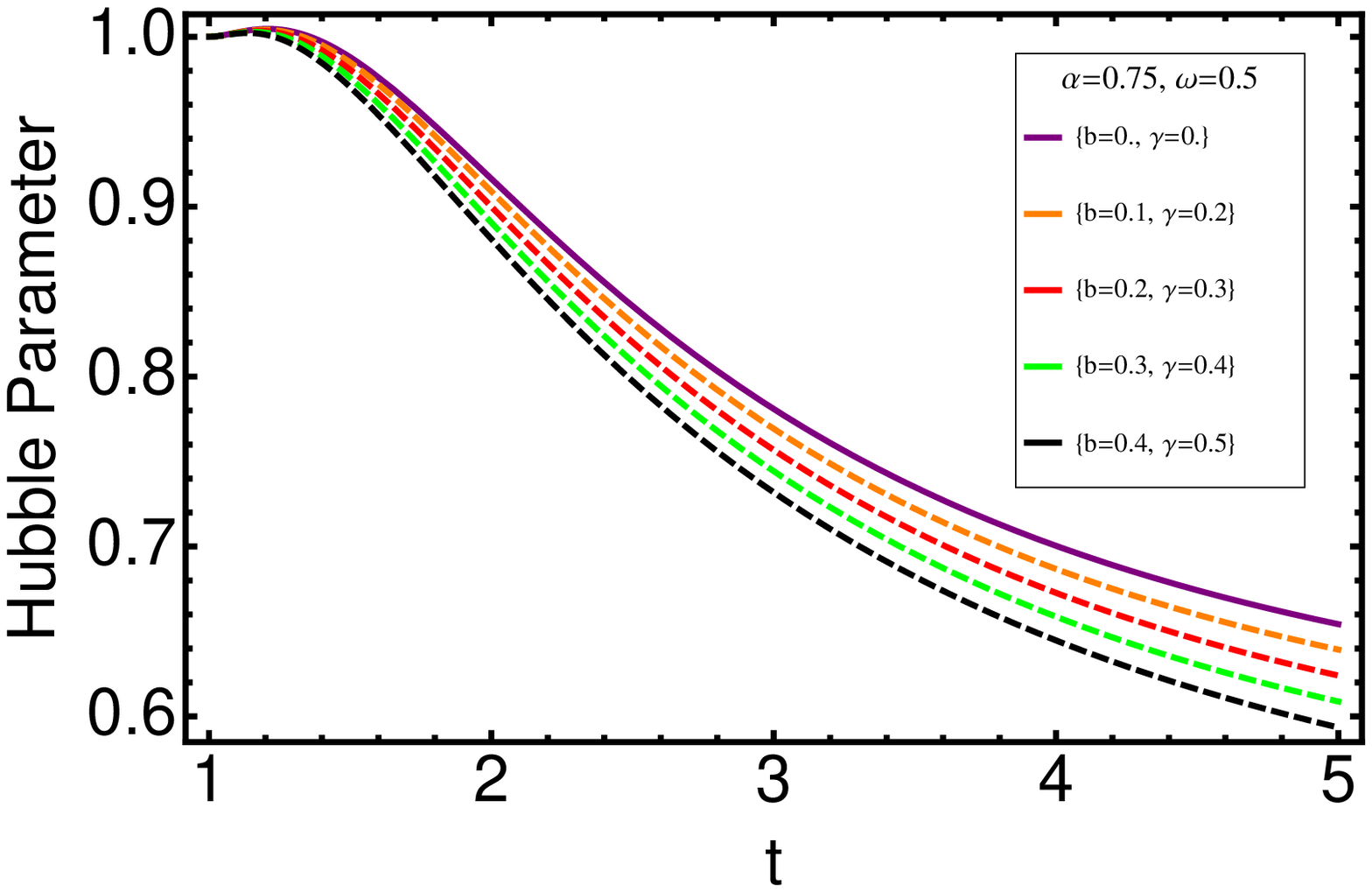}\\
\includegraphics[width=50 mm]{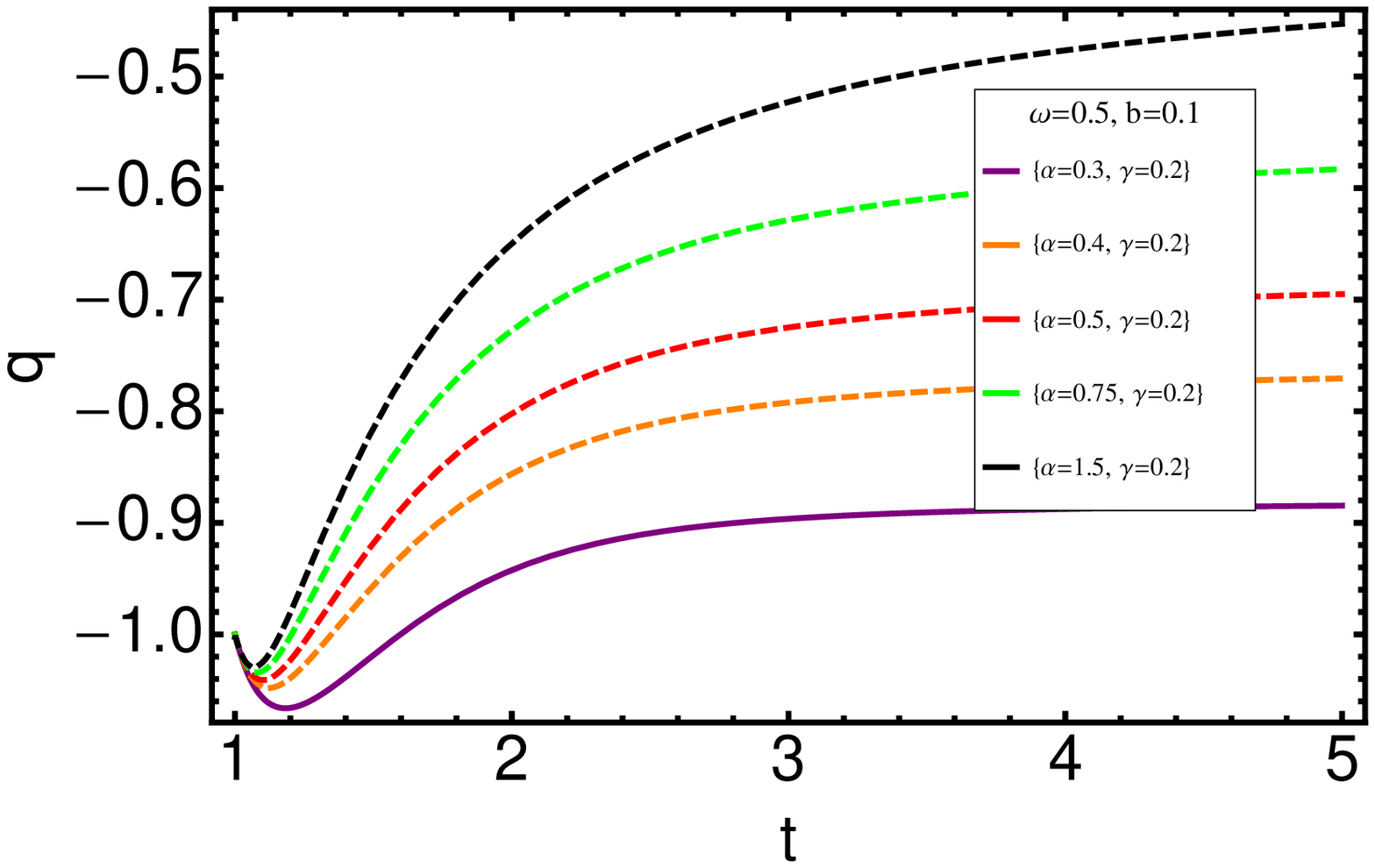} &
\includegraphics[width=50 mm]{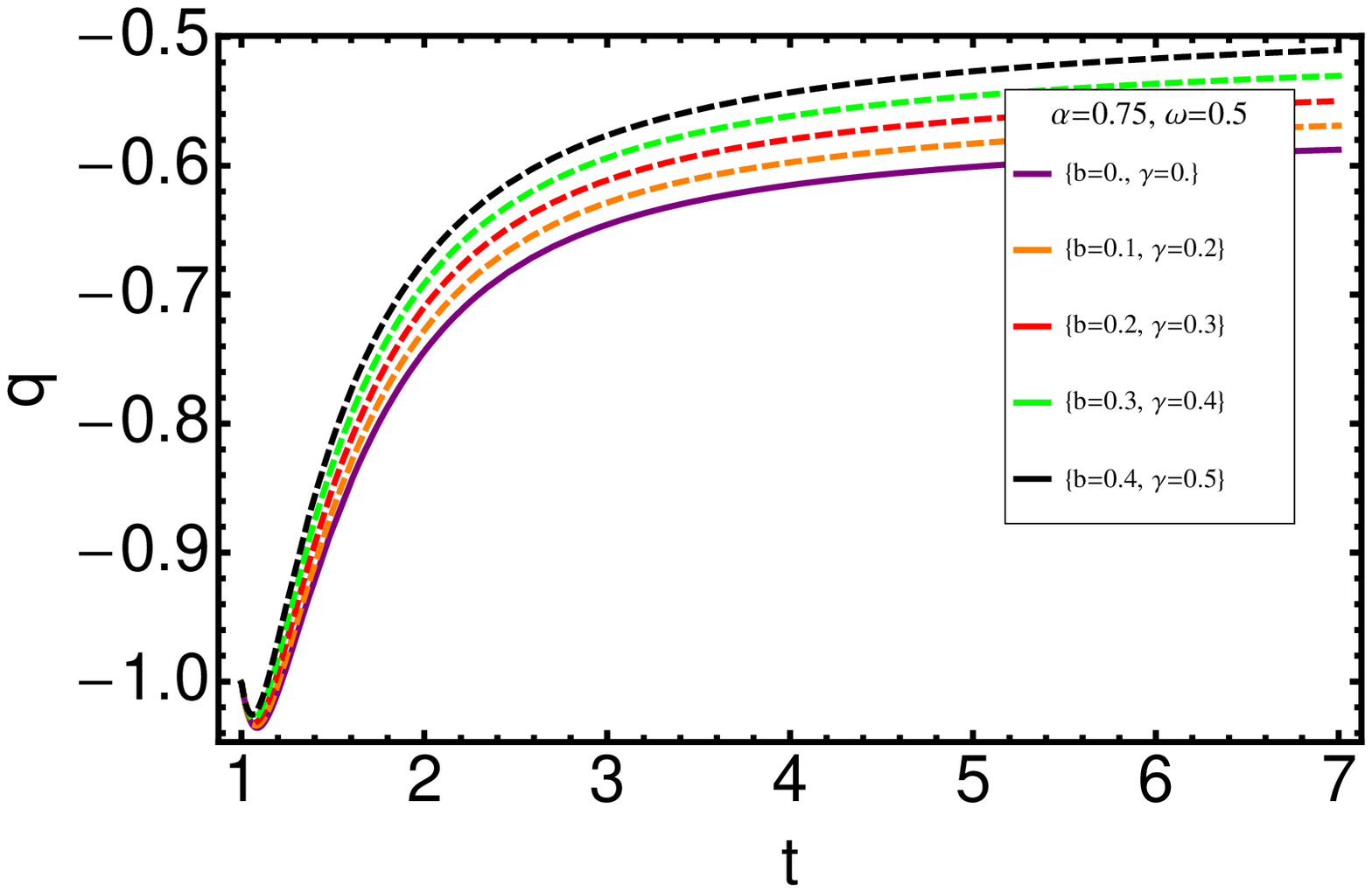}
 \end{array}$
 \end{center}
\caption{Behavior of Hubble parameter $H$ and $q$ against $t$ for $\Lambda=\rho$ and the model 1.}
 \label{fig:1}
\end{figure}

In the plots of Fig. 6 we draw total EoS and Ricci dark energy EoS. In the first plot of Fig. 6 we vary $\alpha$ and find that, at the early universe, value of total EoS decreases by $\alpha$, while at the late time it increases by $\alpha$ to reaches -1. There is a critical point where variation of $\alpha$ is not important. The second plot shows that increasing interaction parameters decrease value of $\omega_{tot}$ as well as $\omega_{R}$ (see last plot). It is illustrated that the Ricci dark energy EoS grows suddenly at the early universe then yields to $\omega_{R}<-1$ at the late time and have phantom-like universe.

\begin{figure}[h!]
 \begin{center}$
 \begin{array}{cccc}
\includegraphics[width=50 mm]{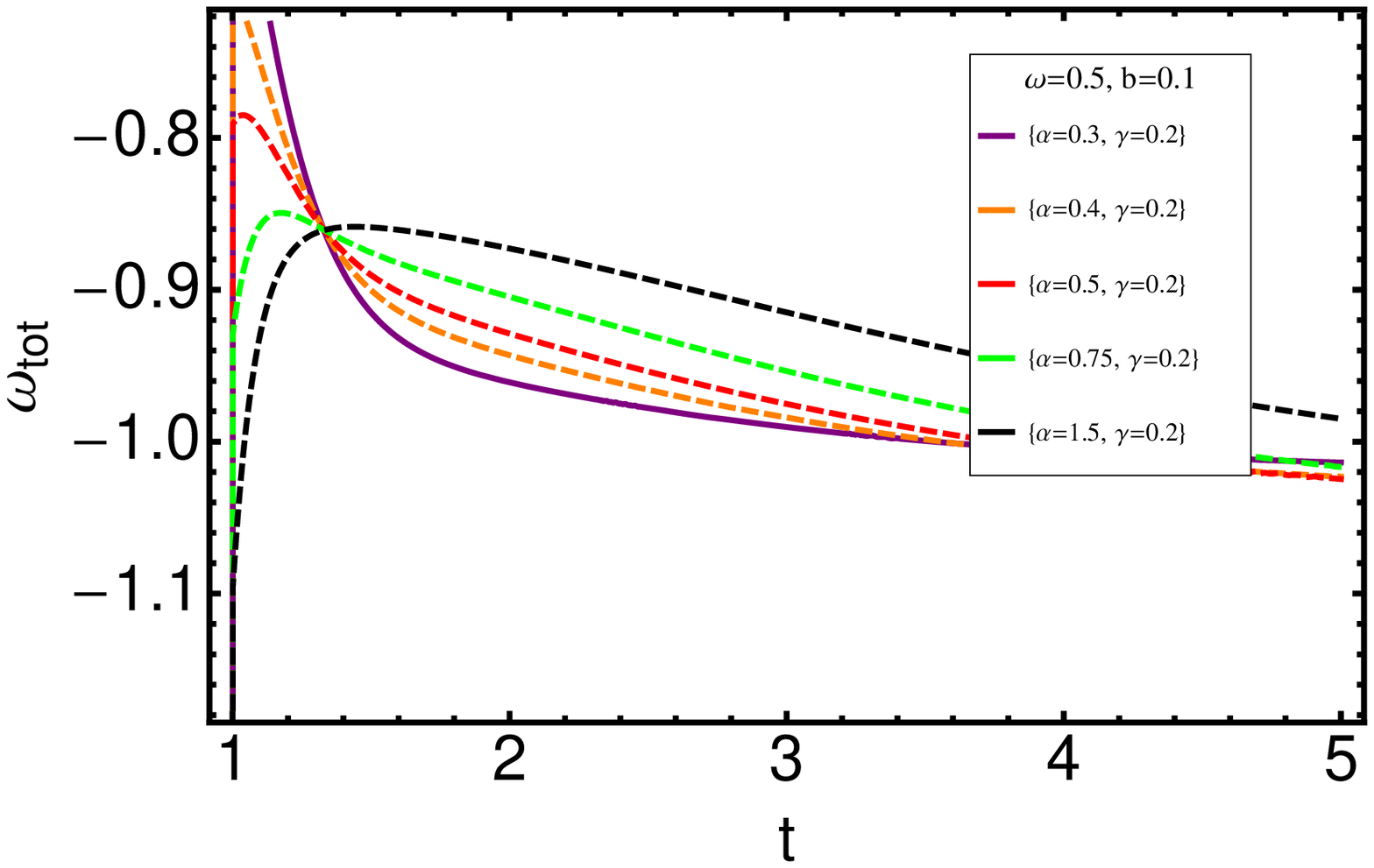} &
\includegraphics[width=50 mm]{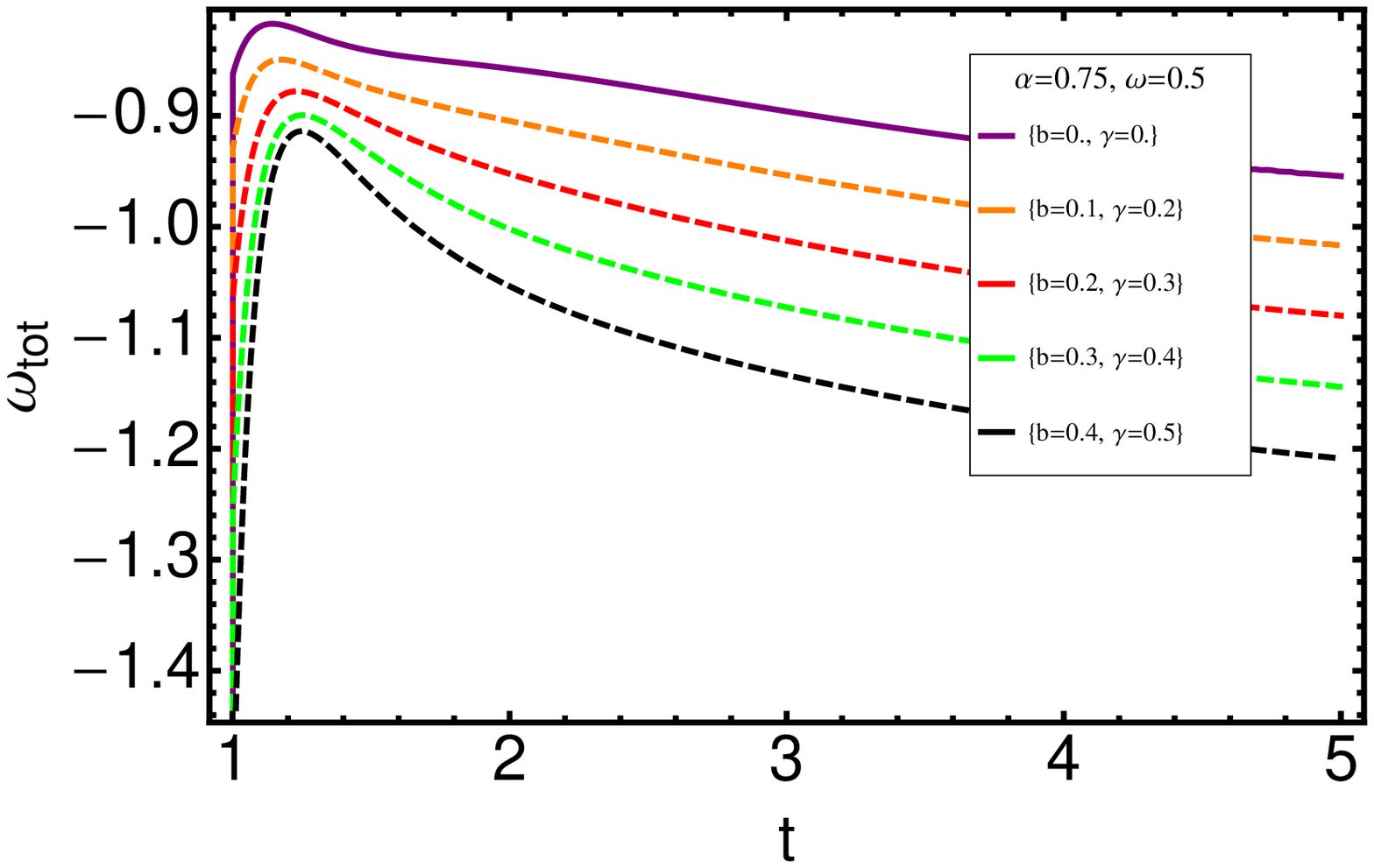} \\
\includegraphics[width=50 mm]{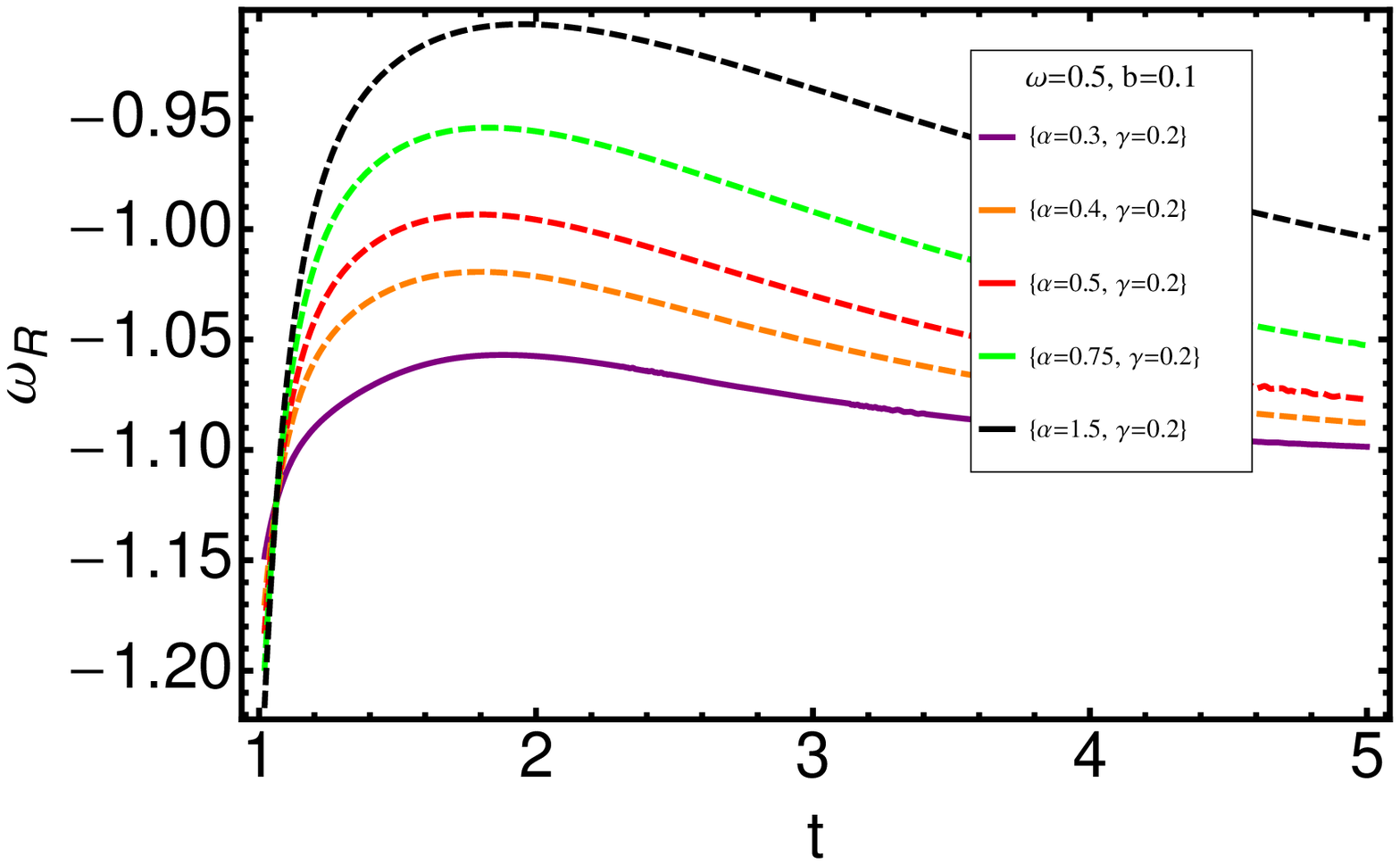} &
\includegraphics[width=50 mm]{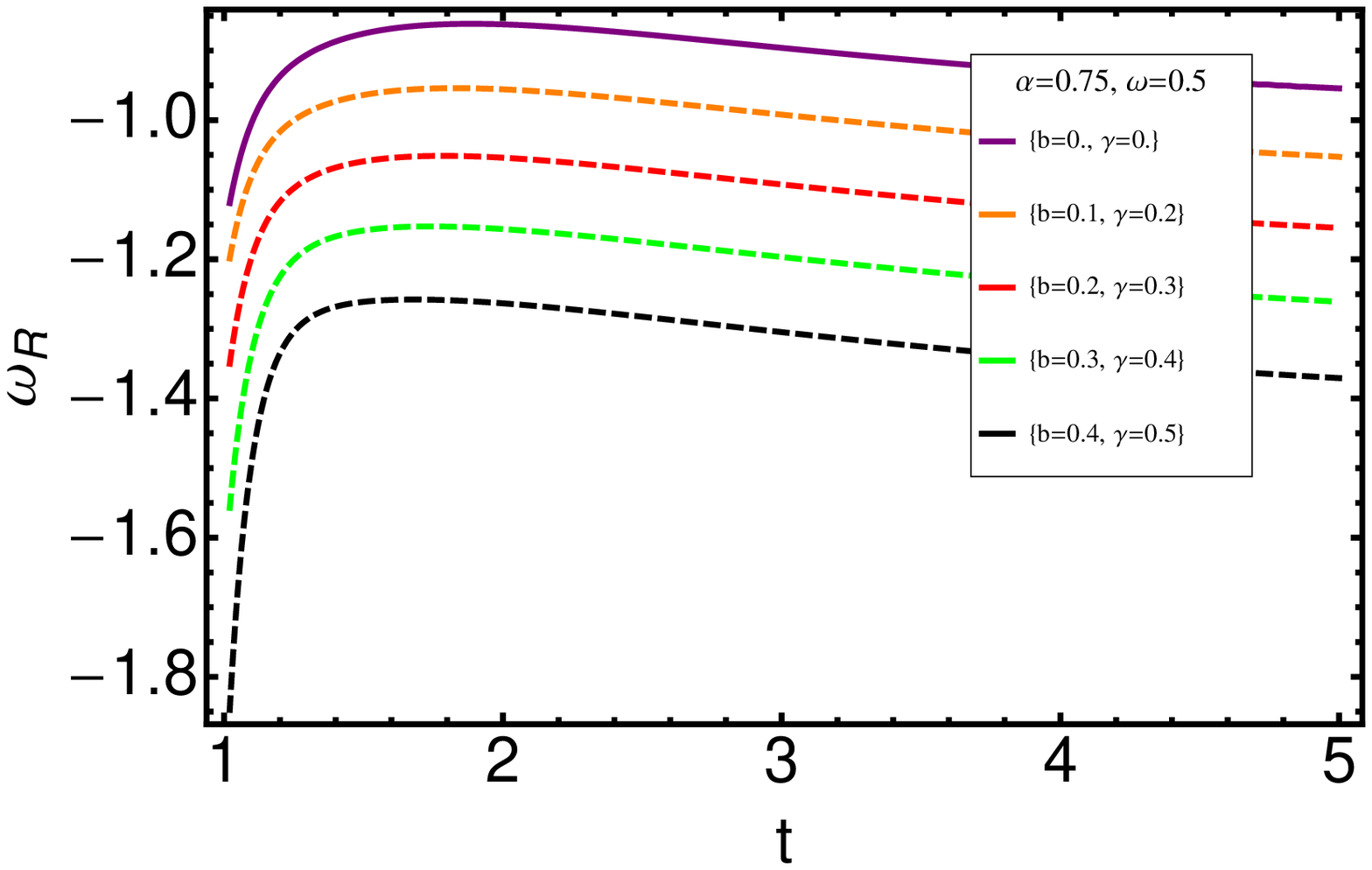}
\end{array}$
 \end{center}
\caption{Behavior of EoS parameter $\omega_{tot}$ and $\omega_{R}$ against $t$ for $\Lambda=\rho$ and the model 1.}
 \label{fig:2}
\end{figure}

\subsection{\large{The model 2}}
As before, in the second model we use the interaction term of the form of Eq. (5). Fig. 7 show the similar behavior (with only infinitesimal changes) of Hubble expansion and the deceleration parameters with model 1, but we can see that variation with interaction parameter is small during the time, specially at the late time there is no important effect. However we can find reasonable behavior comparing with the case of constant $\Lambda$ which tells us that if one can want to consider $\Lambda$ in the dark energy models it is better to choose varying $\Lambda$.\\

\begin{figure}[h!]
 \begin{center}$
 \begin{array}{cccc}
\includegraphics[width=50 mm]{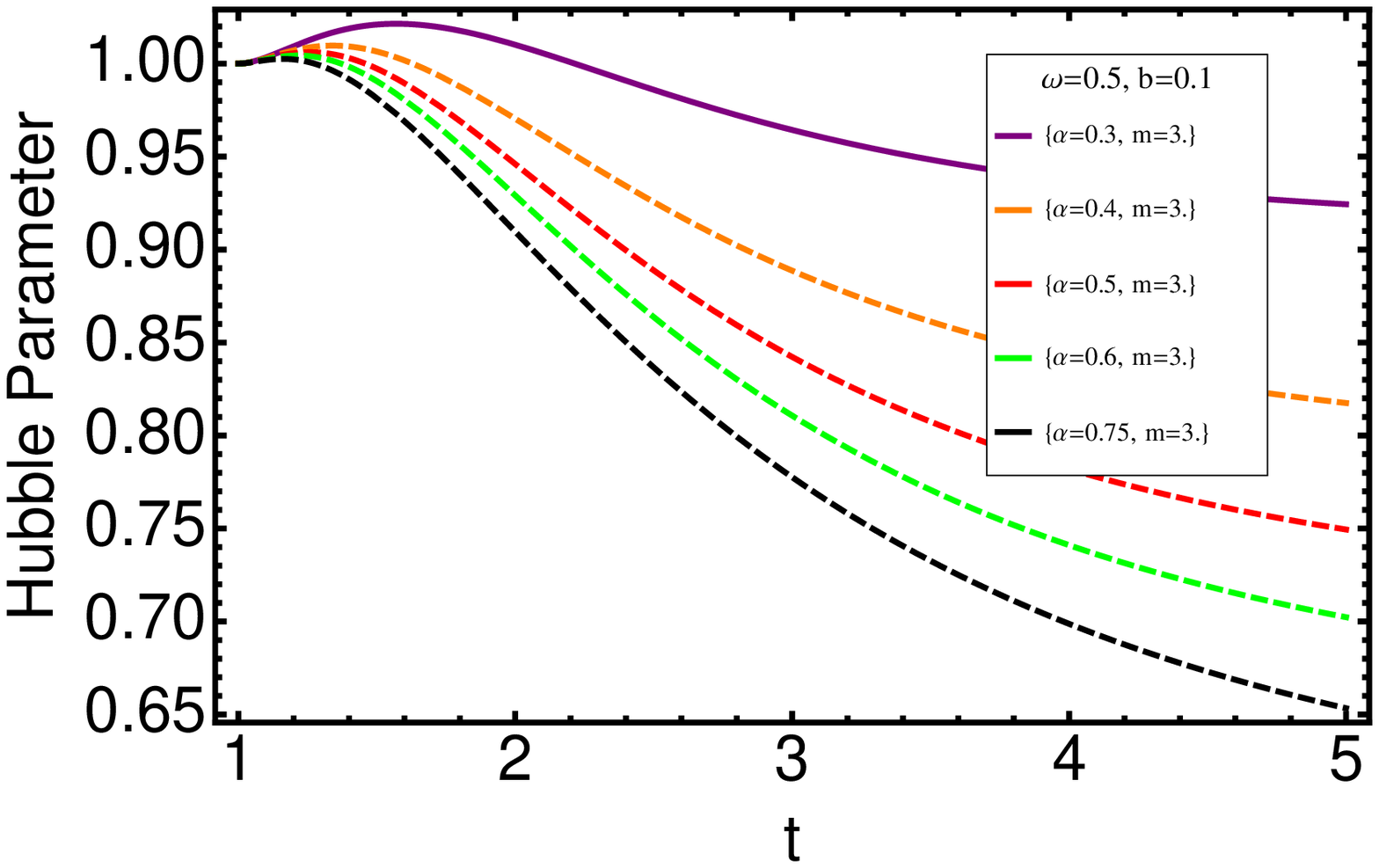} &
\includegraphics[width=50 mm]{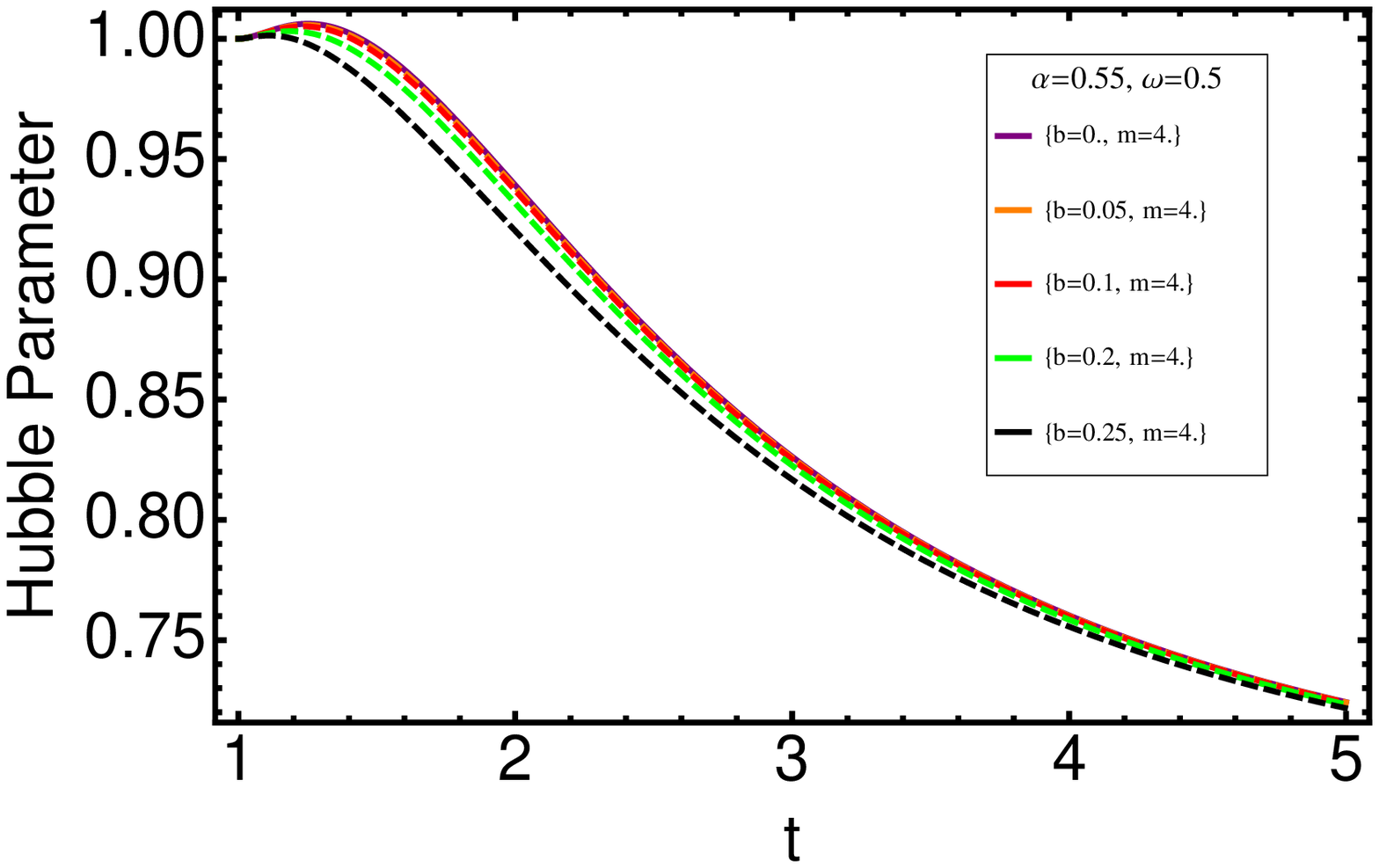}\\
\includegraphics[width=50 mm]{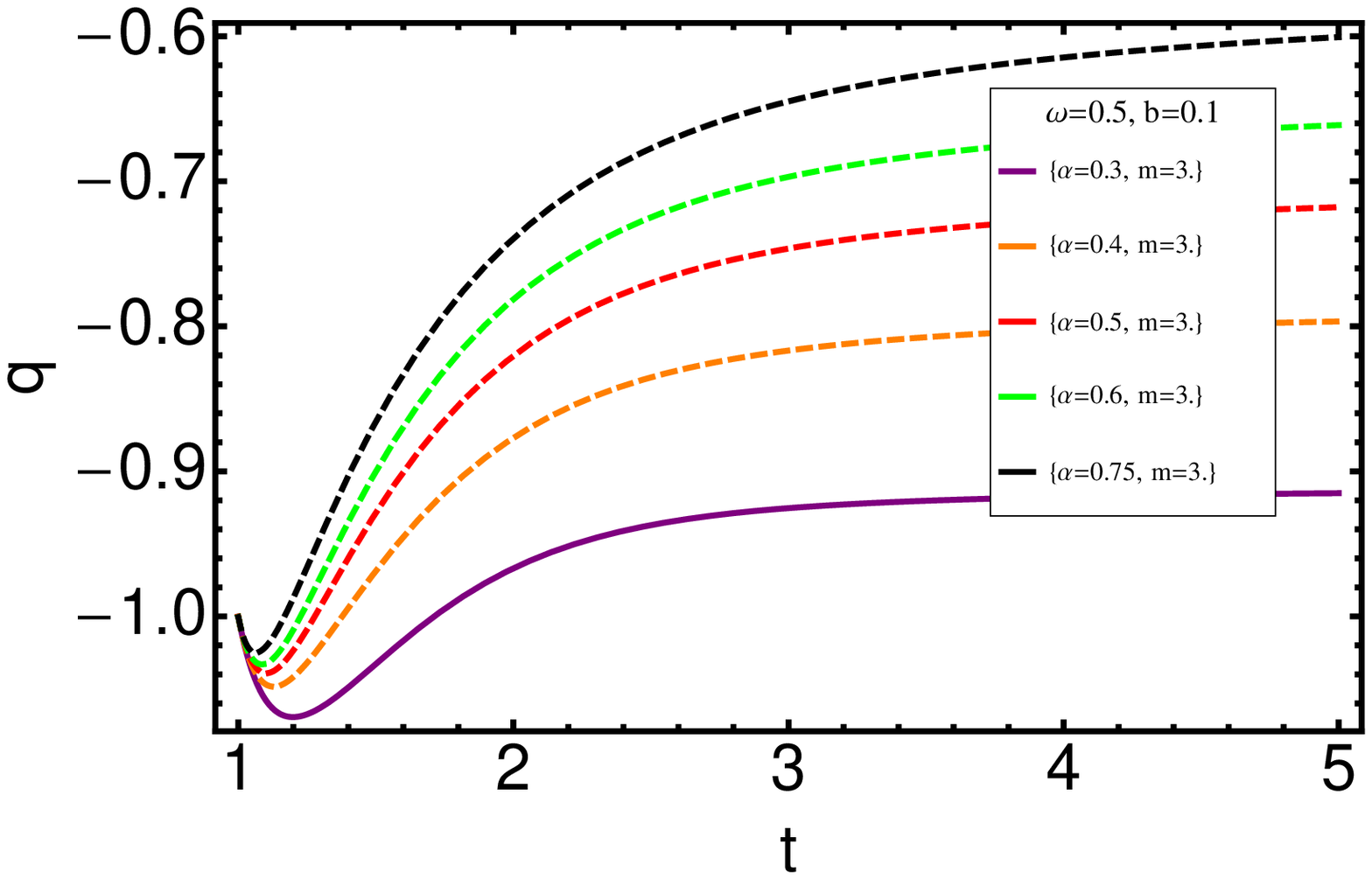} &
\includegraphics[width=50 mm]{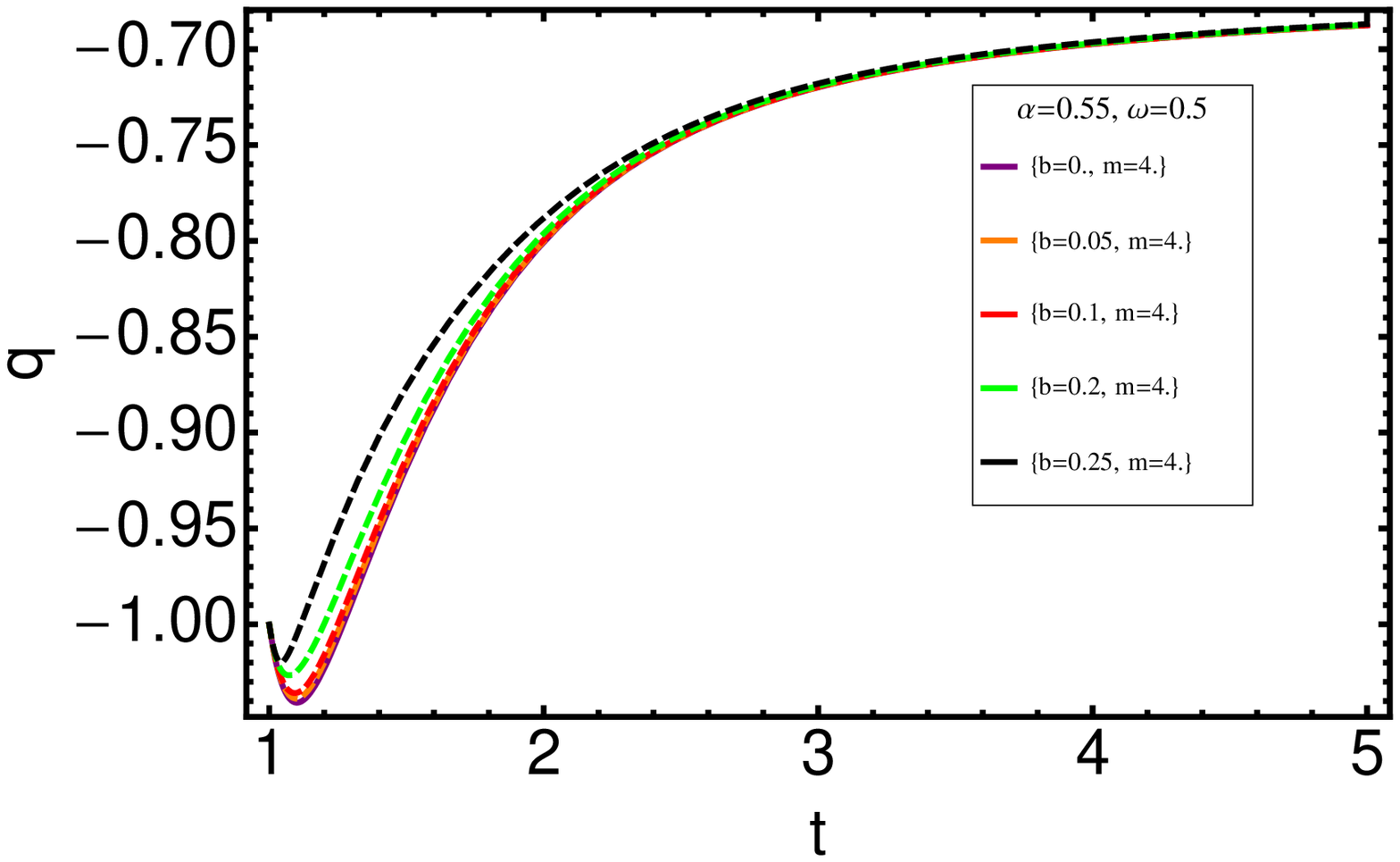}
 \end{array}$
 \end{center}
\caption{Behavior of Hubble parameter $H$ and $q$ against $t$ for $\Lambda=\rho$ and the model 2.}
 \label{fig:3}
\end{figure}

\begin{figure}[h!]
 \begin{center}$
 \begin{array}{cccc}
\includegraphics[width=50 mm]{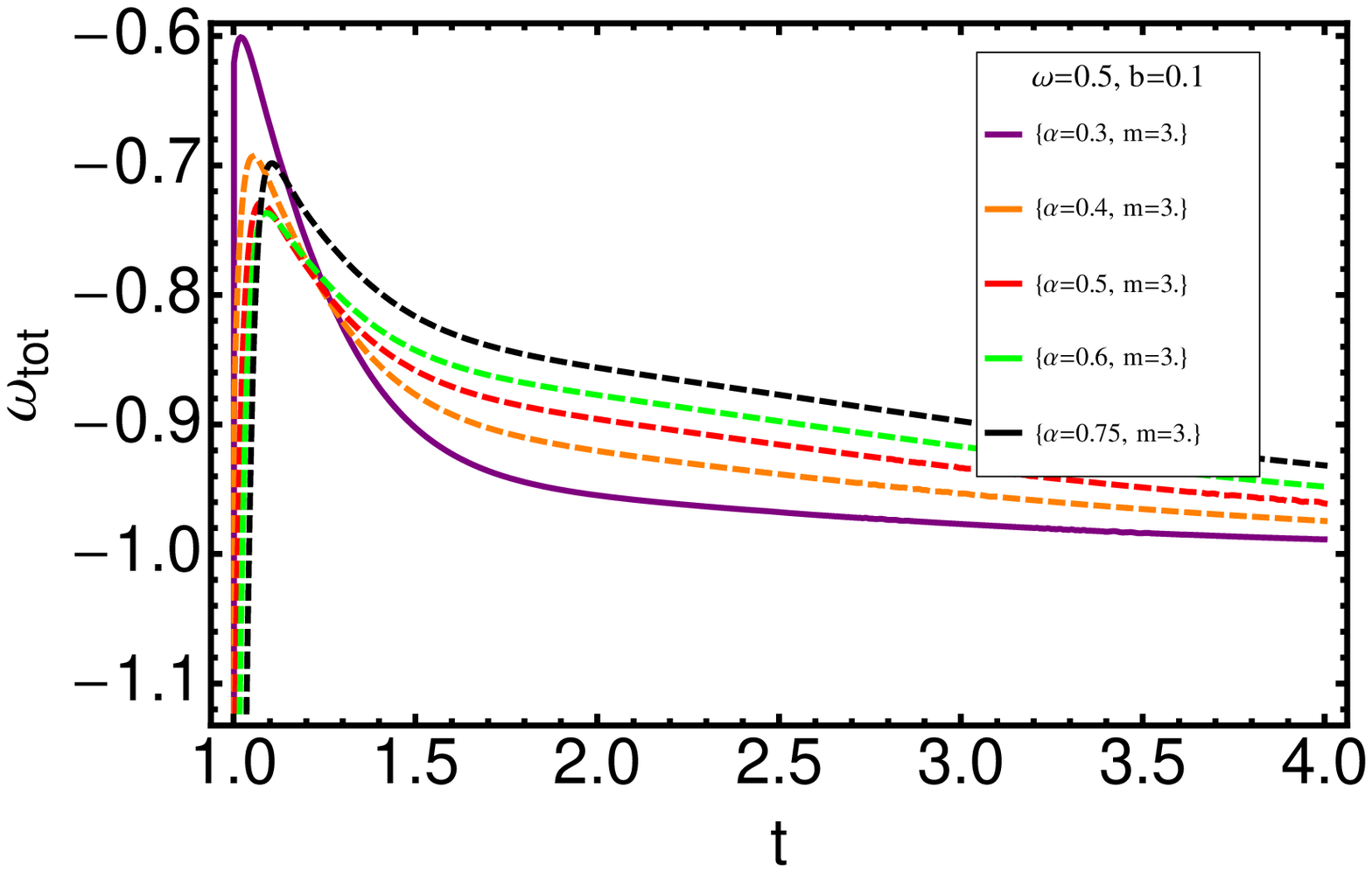} &
\includegraphics[width=50 mm]{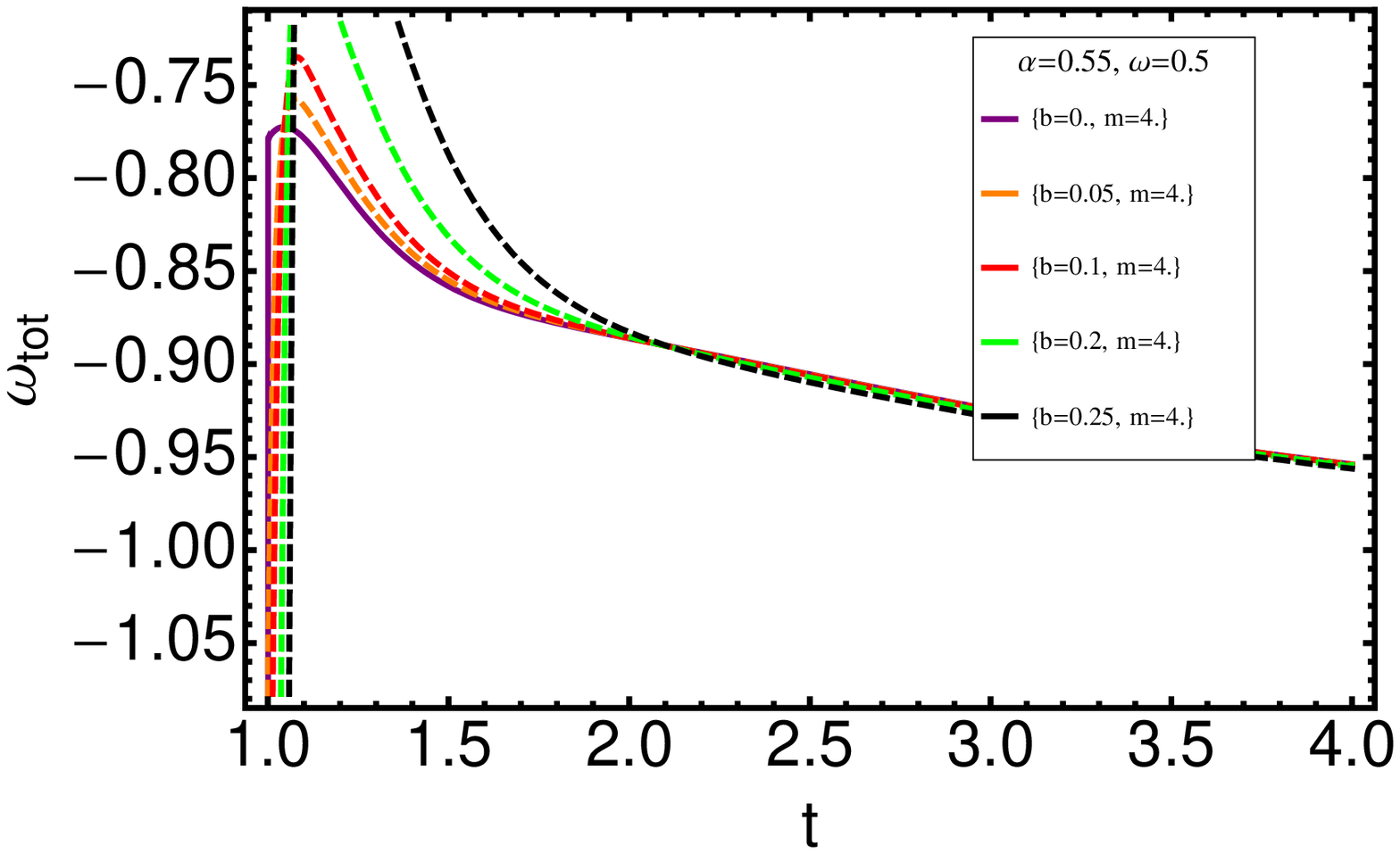}\\
\includegraphics[width=50 mm]{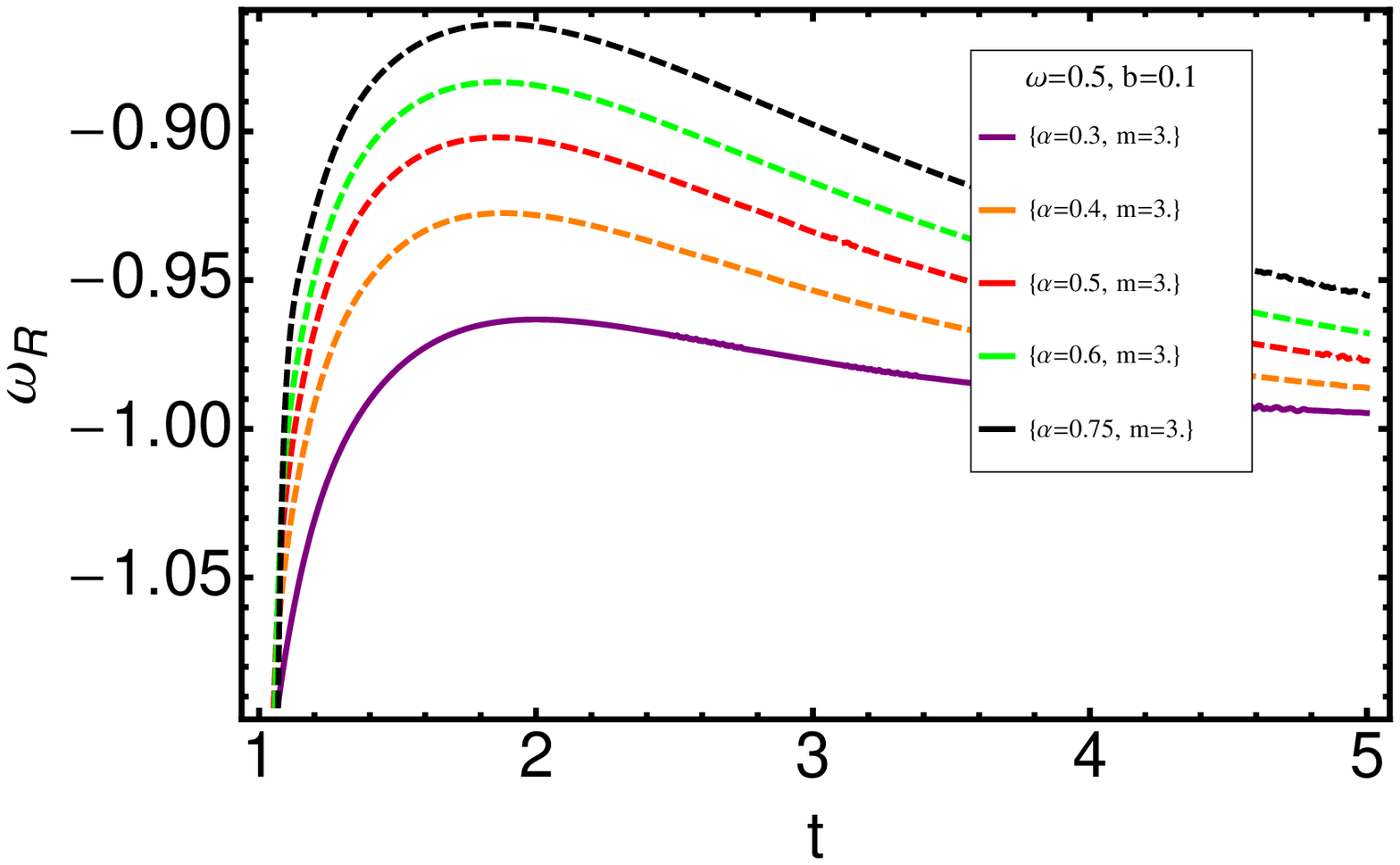} &
\includegraphics[width=50 mm]{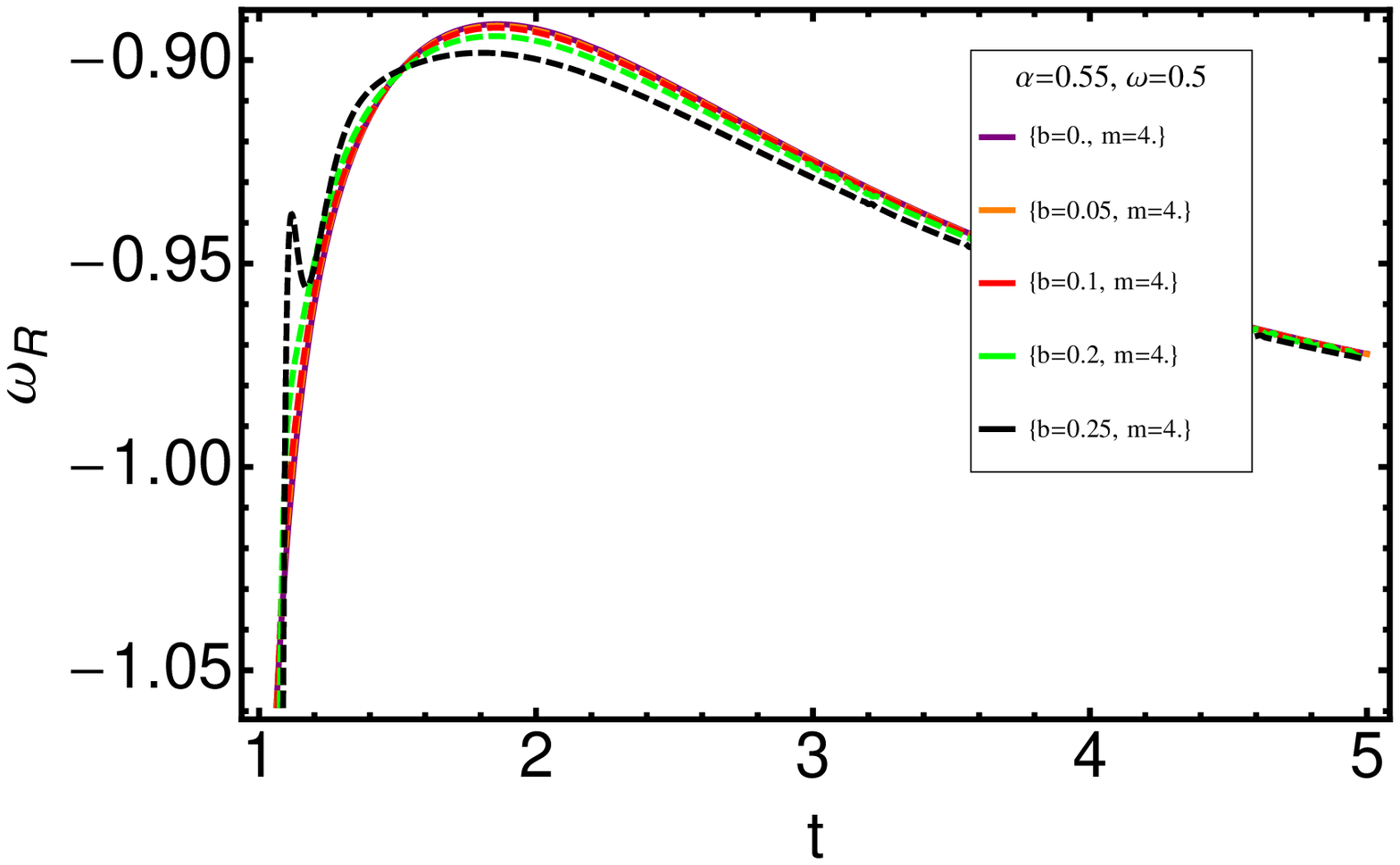}
\end{array}$
 \end{center}
\caption{Behavior of EoS parameter $\omega_{tot}$ and $\omega_{R}$ against $t$ for $\Lambda=\rho$ and the model 2.}
 \label{fig:4}
\end{figure}

On the other hand plots of the Fig. 8 represent EoS parameters. We can see that total EoS parameter grows suddenly at the early universe to reach a maximum value, then reduced to reach a constant value ($\omega_{tot}\rightarrow-1$) at the late time. We also find that increasing $\alpha$ increases value of $\omega_{tot}$ at the late time. As before, there is no difference between various values of interaction parameter $b$ on the $\omega_{tot}$ at the late epoch. Time evolution of Ricci dark energy EoS shows that $\omega_{R}<-1$ at the early universe but $\omega_{R}>-1$ at the late time. It means that the universe is phantom like at the beginning and translates to quintessence like at present. Therefore, it is completely different with the model 1. Hence this model is inconsistent with observational data.

\section{\large{The case of ($t, H, \rho$)-dependent $\Lambda$ }}
In this section we assume $\Lambda$ given by the relation (7). It has indeed three terms. The first term is inverse of squared time, which may be negligible at the late time. The second term is squared Hubble parameter which is a constant at the late time. Finally the last term is total density multiple by exponential of $-tH$. The last term also may negligible at the late time. Therefore, we can see that the value of $\Lambda$ at the late time is a negative constant which is characteristics of anti-de Sitter space-time.
\subsection{\large{The model 1}}
Again, we choose interaction term (4) to have numerical analysis. Fig. 9 tells us that the Hubble expansion parameter is totally increasing function of time for small values of $\alpha$ and it is not depend on interaction parameters. In order to have reasonable behavior of Hubble expansion parameter we should choose larger value of $\alpha$ ($\geq0.6$). Similar results obtained by analyzing the deceleration parameter. For the appropriate choice of $\alpha$ the deceleration parameter yields to -1 at the late time in agreement with $\Lambda$CDM model.\\
EoS parameters of this model investigated by using plots of Fig. 10. As previous in this model we can obtain $\omega_{tot}\rightarrow-1$ for $\alpha\geq0.6$. Also increasing the interaction parameters increases value of $\omega_{tot}$ at the late time but decreases one at the early universe. Also, Ricci dark energy EoS indicates that every when we have phantom like universe.\\

\begin{figure}[h!]
 \begin{center}$
 \begin{array}{cccc}
\includegraphics[width=50 mm]{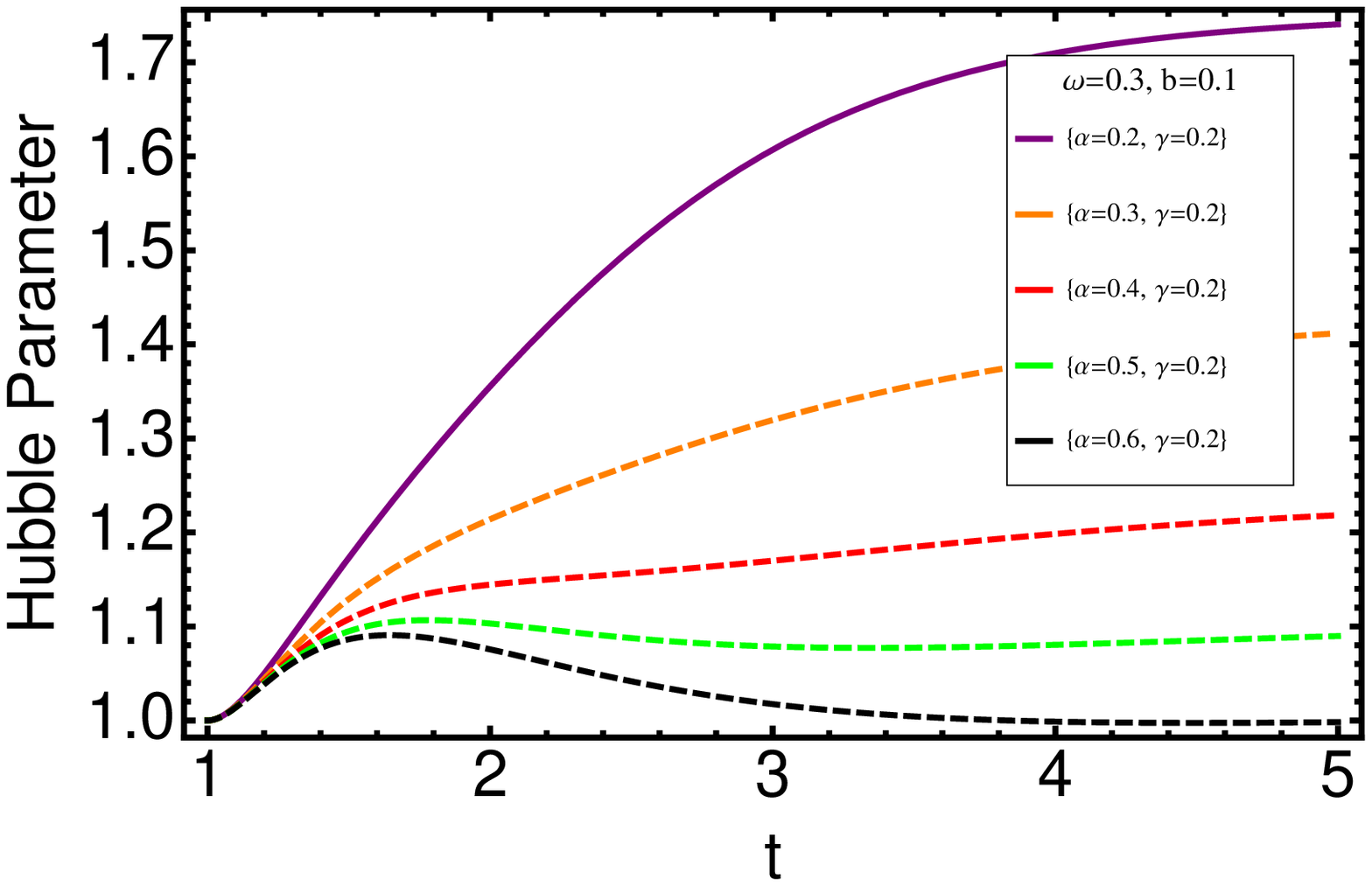} &
\includegraphics[width=50 mm]{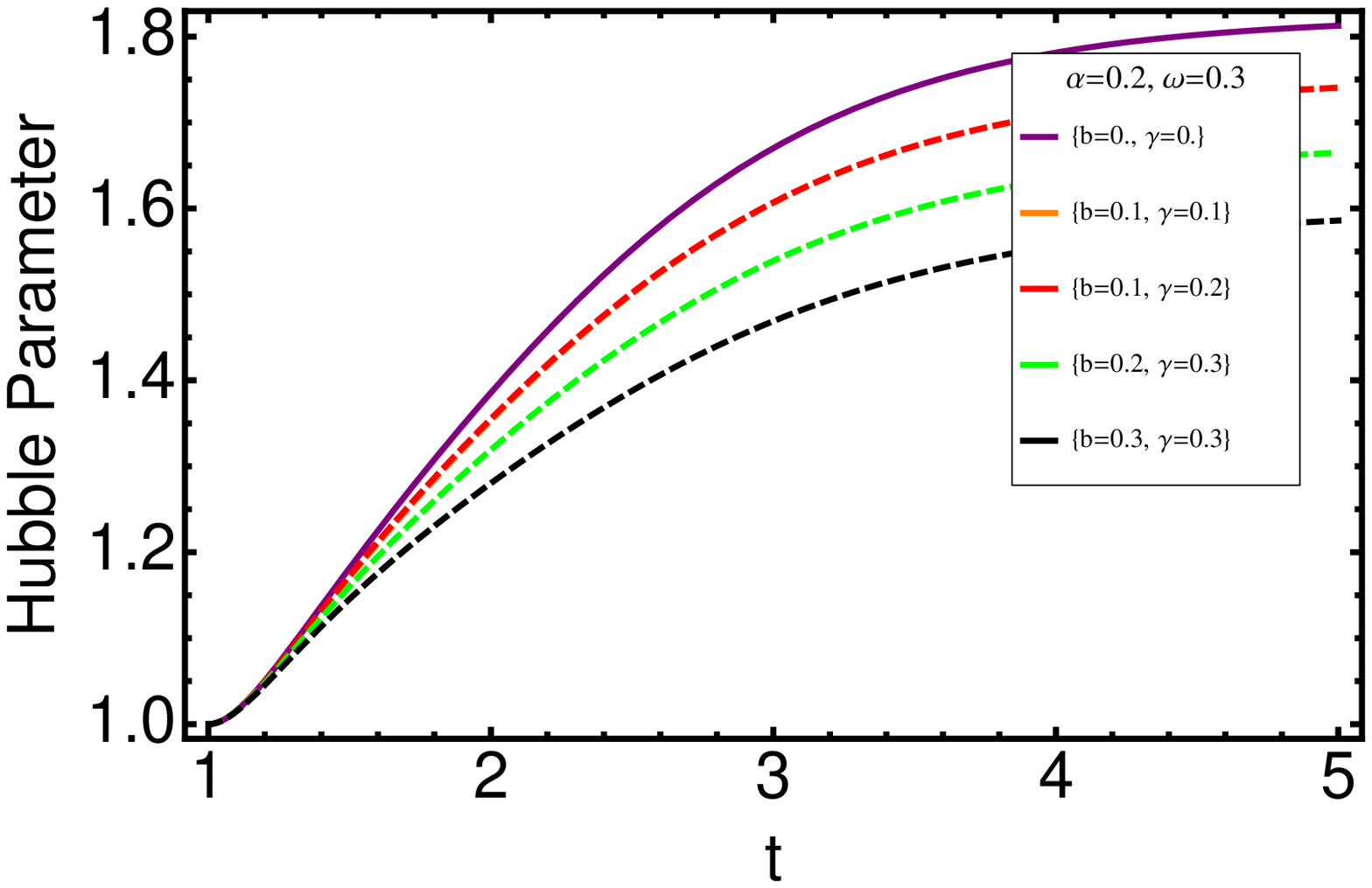}\\
\includegraphics[width=50 mm]{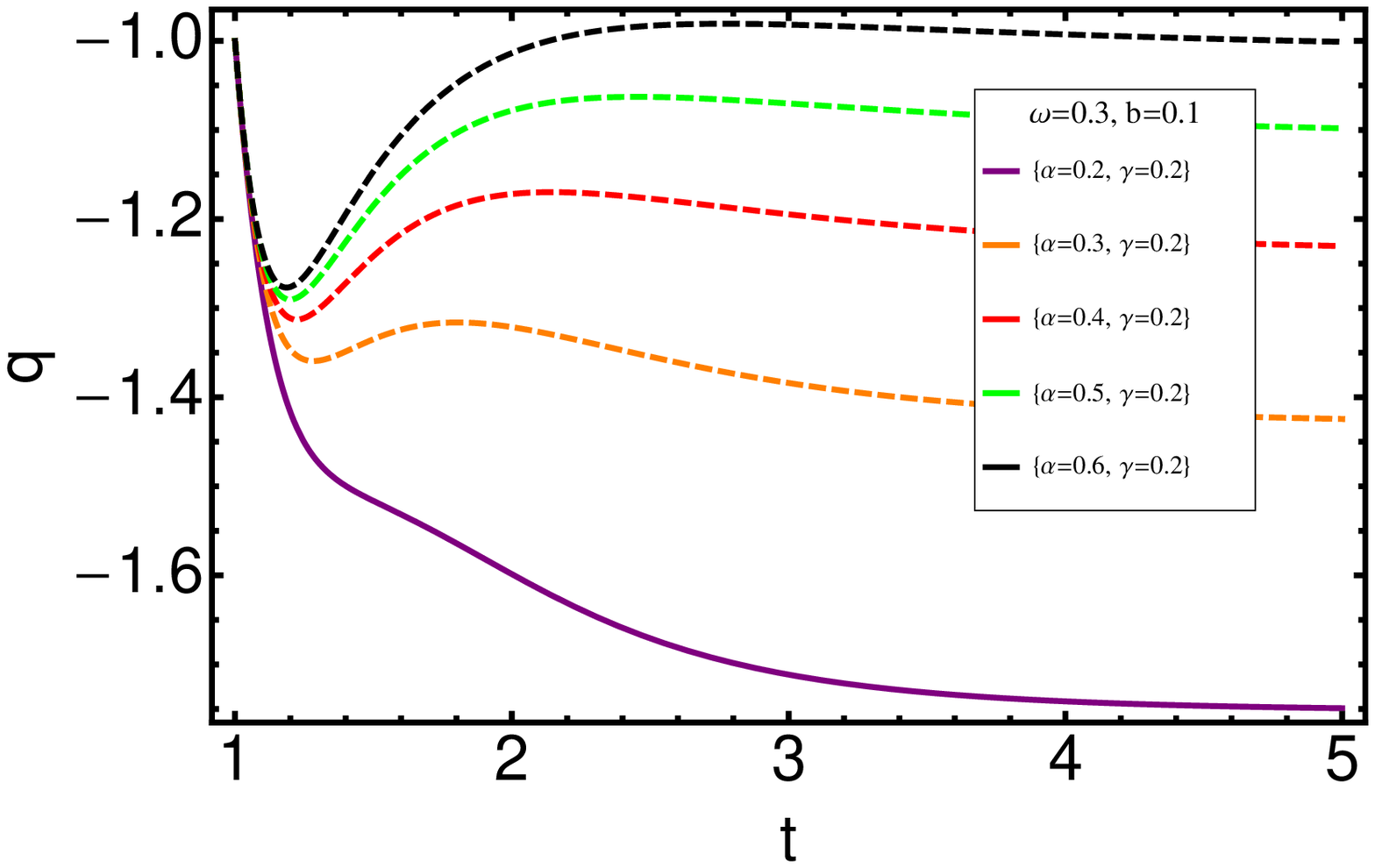} &
\includegraphics[width=50 mm]{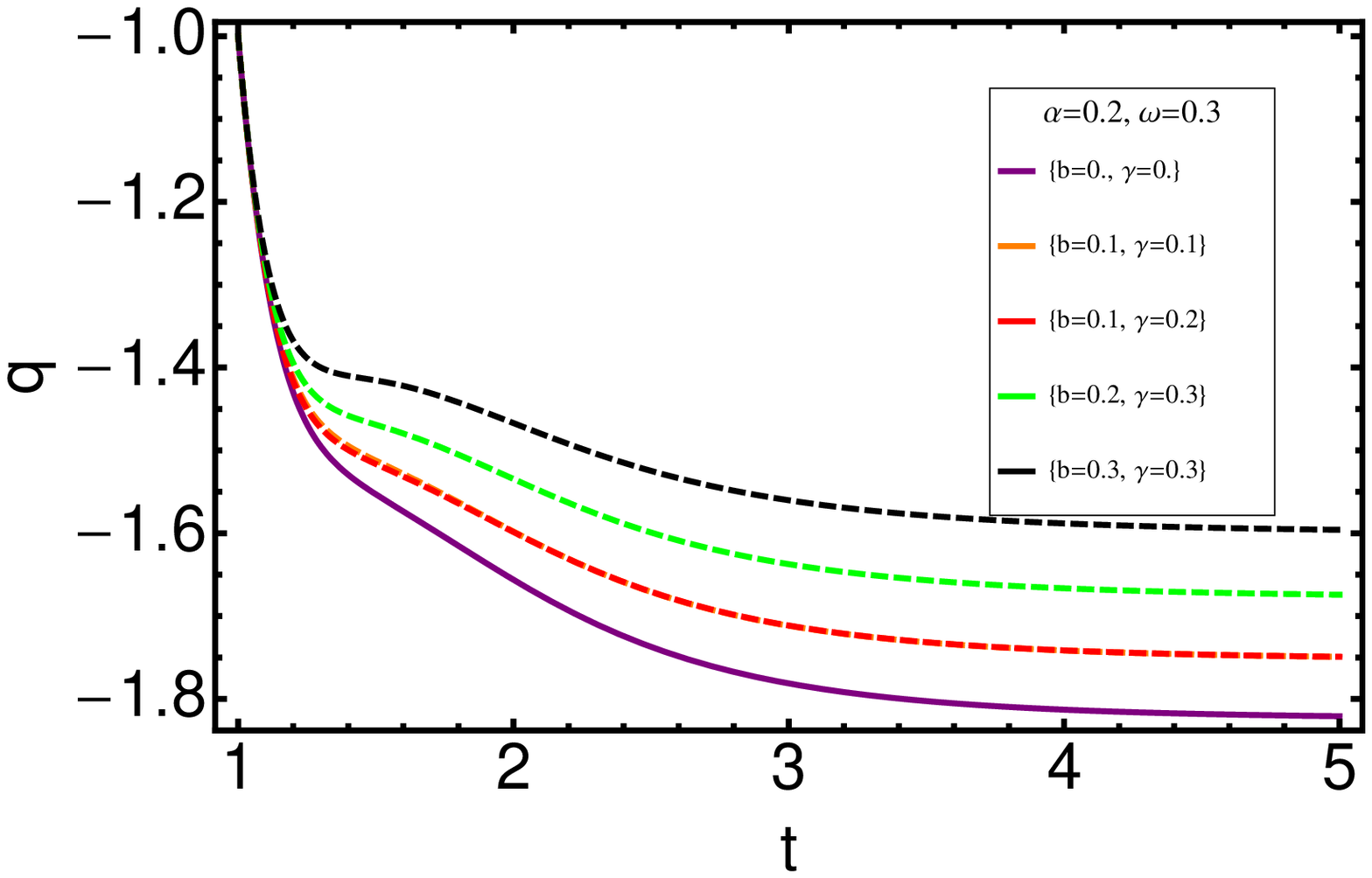}
 \end{array}$
 \end{center}
\caption{Behavior of Hubble parameter $H$ and $q$ against $t$ for ($t, H, \rho$)-dependent $\Lambda$ and the model 1.}
 \label{fig:5}
\end{figure}

\begin{figure}[h!]
 \begin{center}$
 \begin{array}{cccc}
\includegraphics[width=50 mm]{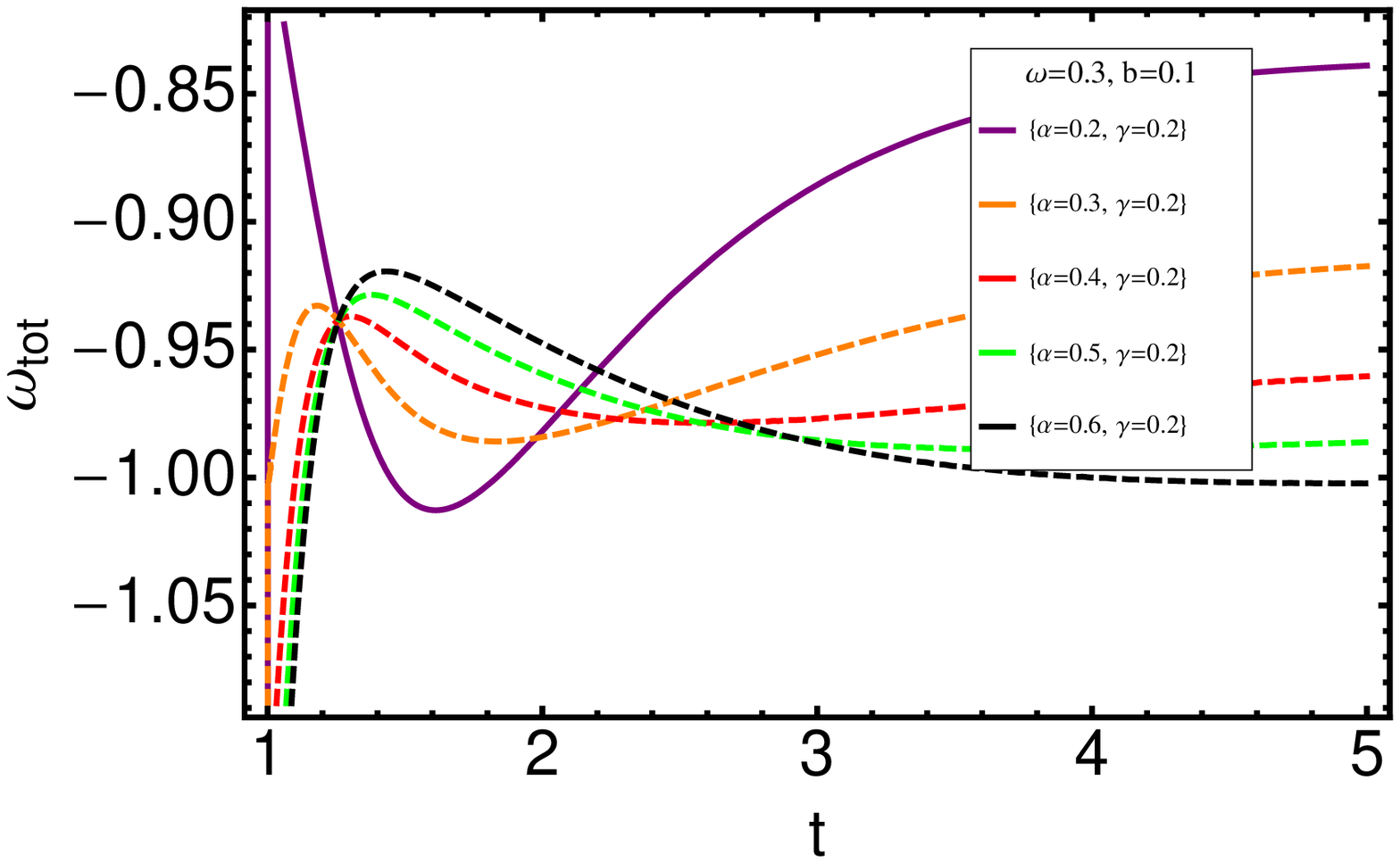} &
\includegraphics[width=50 mm]{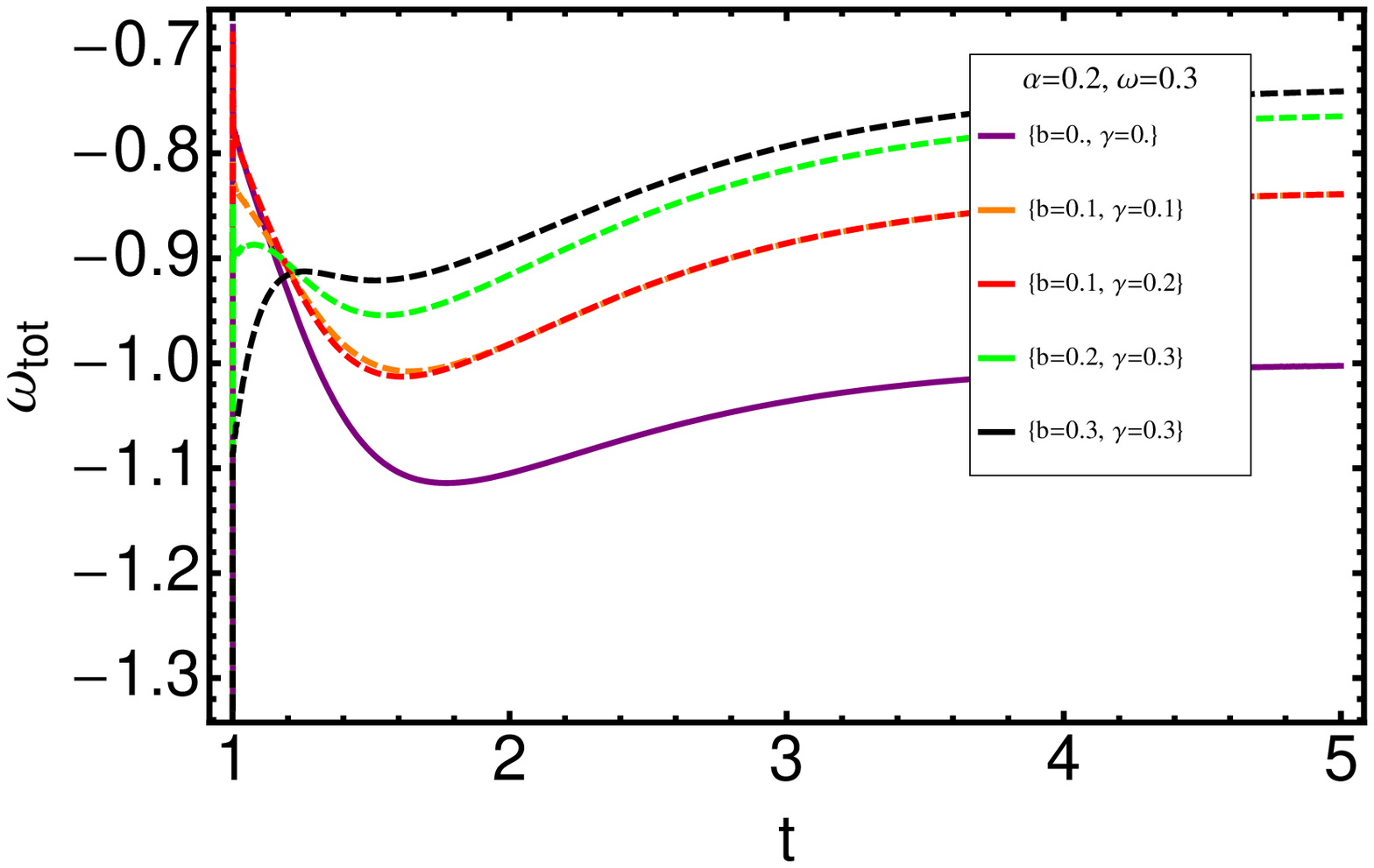} \\
\includegraphics[width=50 mm]{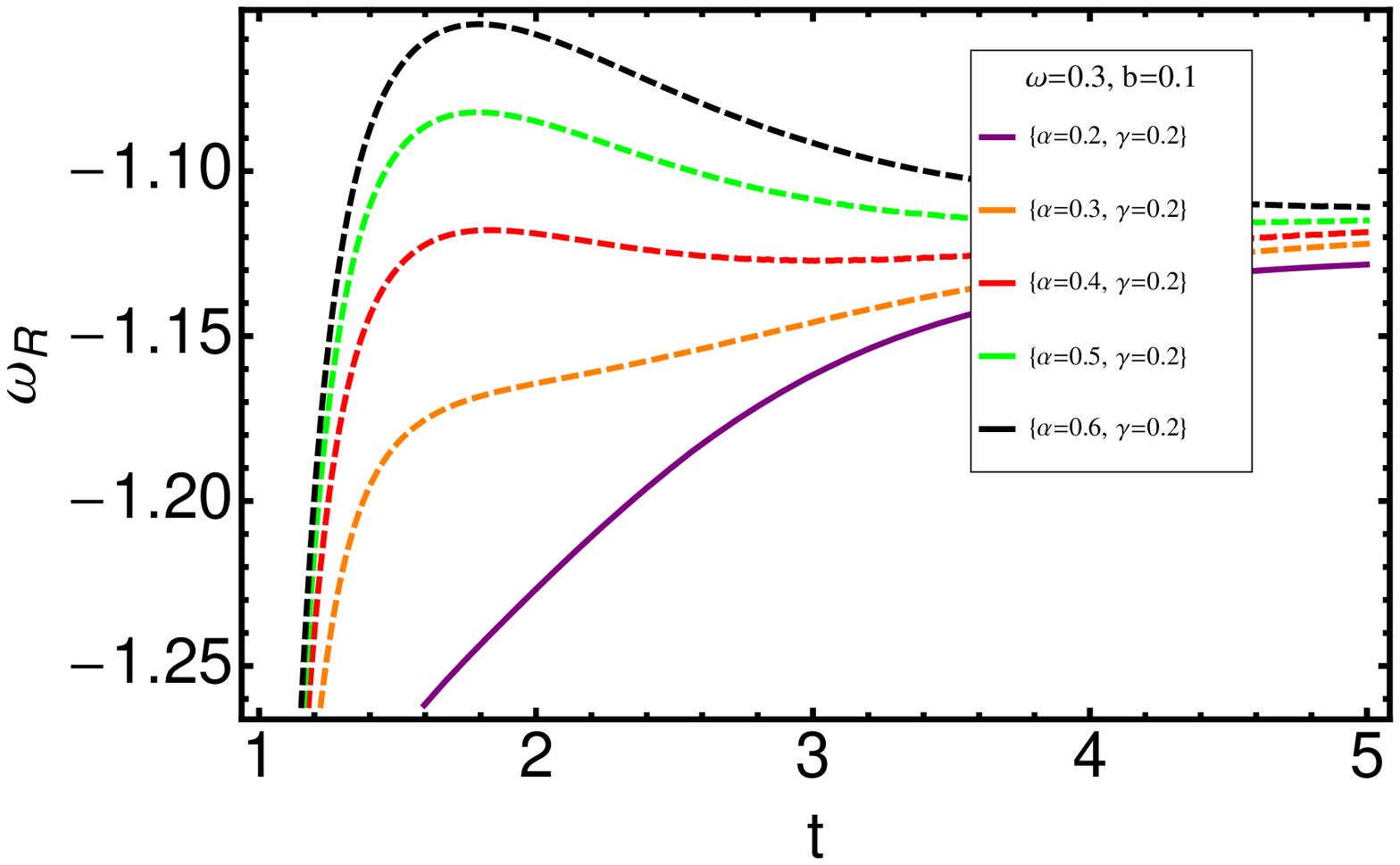} &
\includegraphics[width=50 mm]{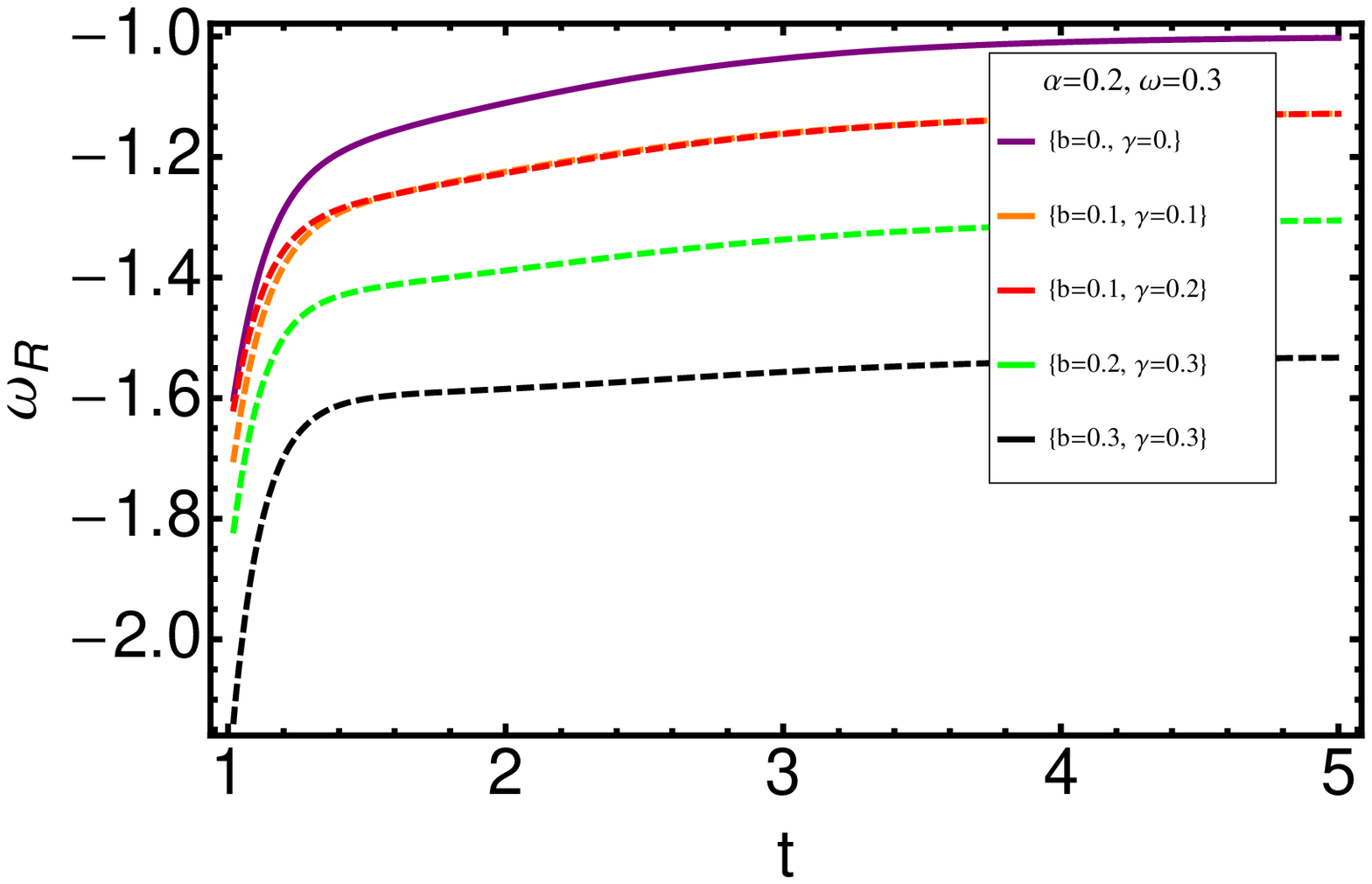}
\end{array}$
 \end{center}
\caption{Behavior of EoS parameter $\omega_{tot}$ and $\omega_{R}$ against $t$ for ($t, H, \rho$)-dependent $\Lambda$ and the model 1.}
 \label{fig:6}
\end{figure}

\subsection{\large{The model 2}}
In the model 2 we use interaction term (5) and obtain behavior of Hubble, deceleration and EoS parameters numerically which illustrated in Figs. 11 and 12. Fig. 11 shows that variation of $m$ is not important in evolution of $H$ and $q$. Also, Fig. 12 indicates that the early universe is phantom like while the late universe is in quintessence like.

\begin{figure}[h!]
 \begin{center}$
 \begin{array}{cccc}
\includegraphics[width=50 mm]{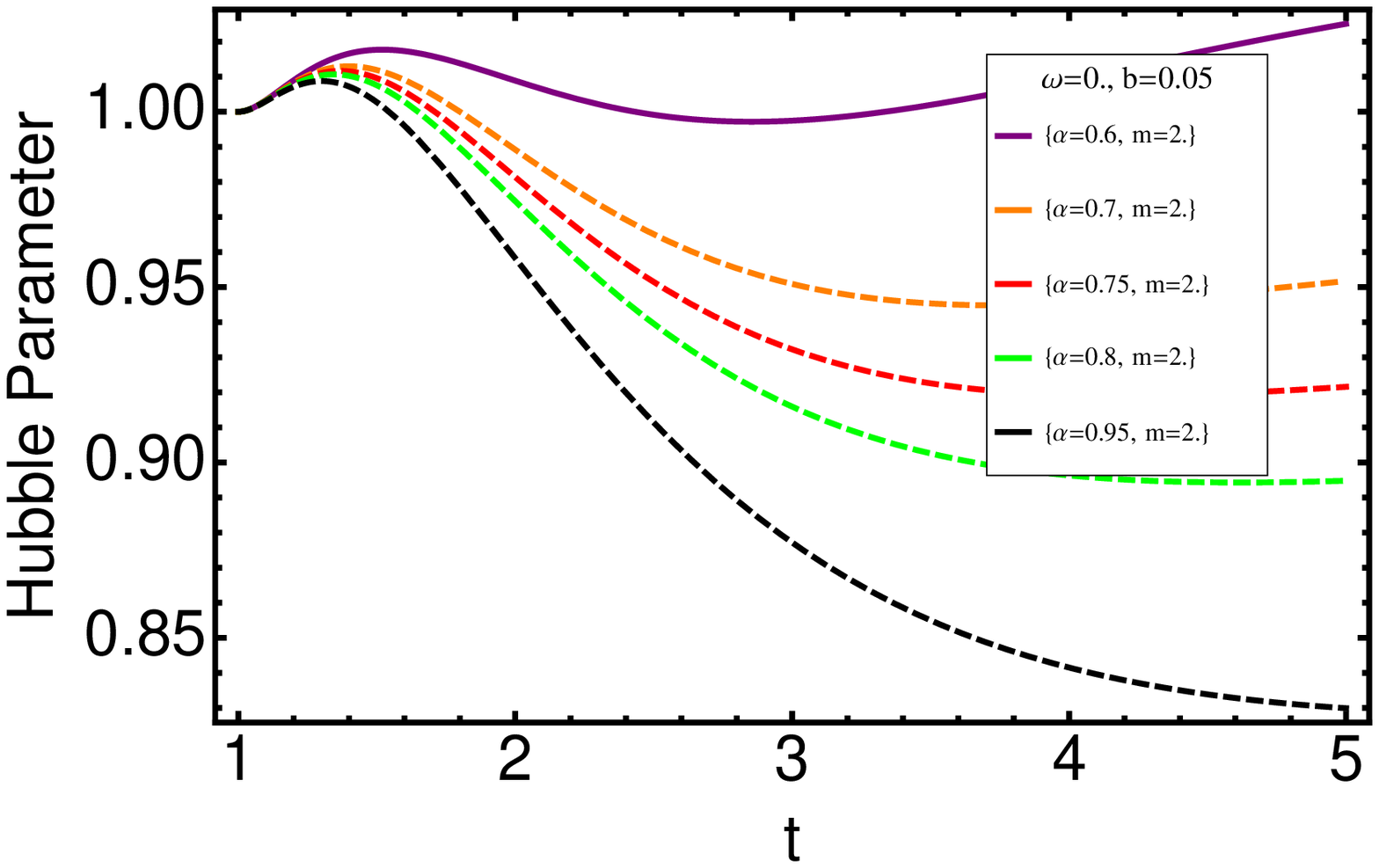} &
\includegraphics[width=50 mm]{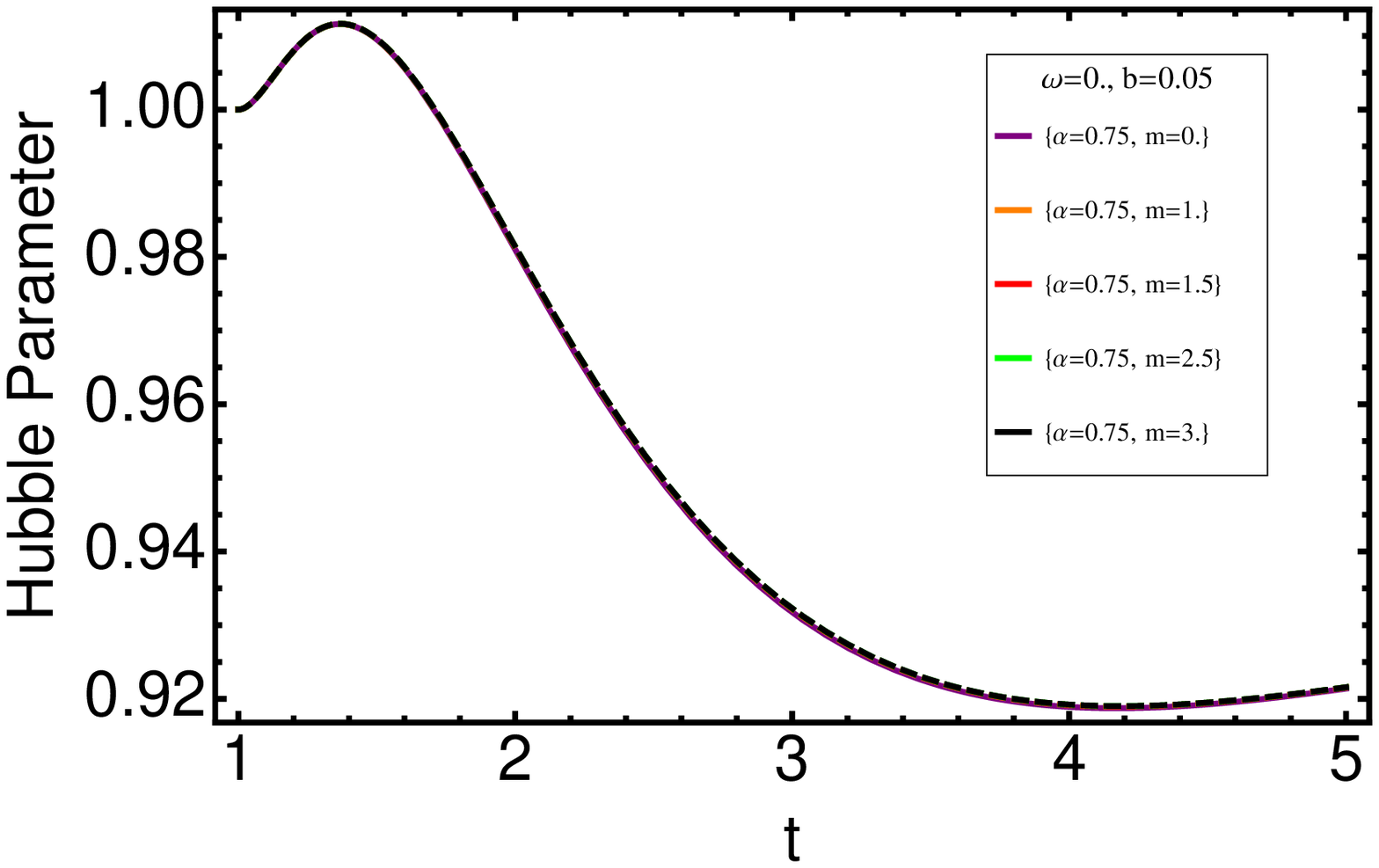}\\
\includegraphics[width=50 mm]{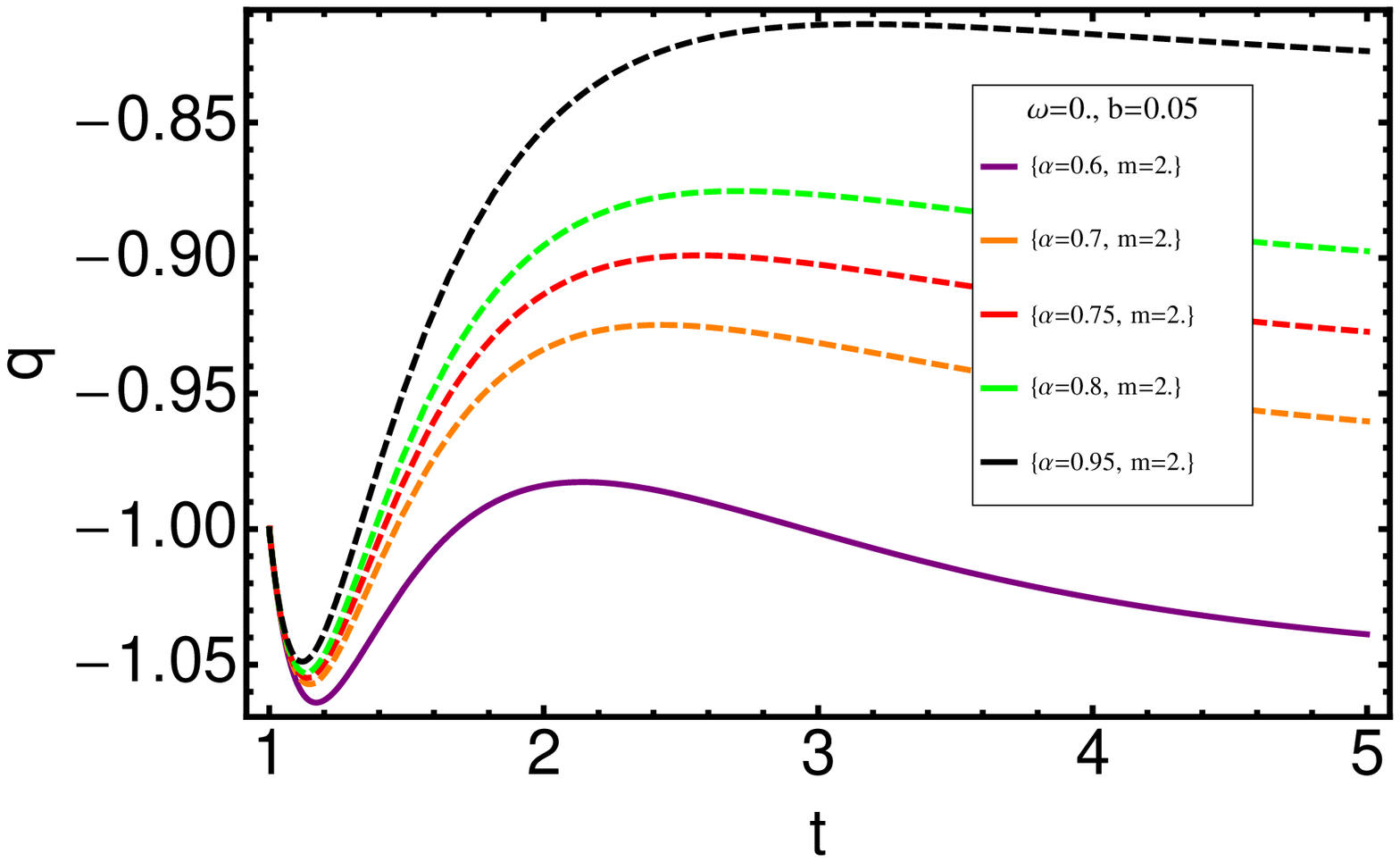}&
\includegraphics[width=50 mm]{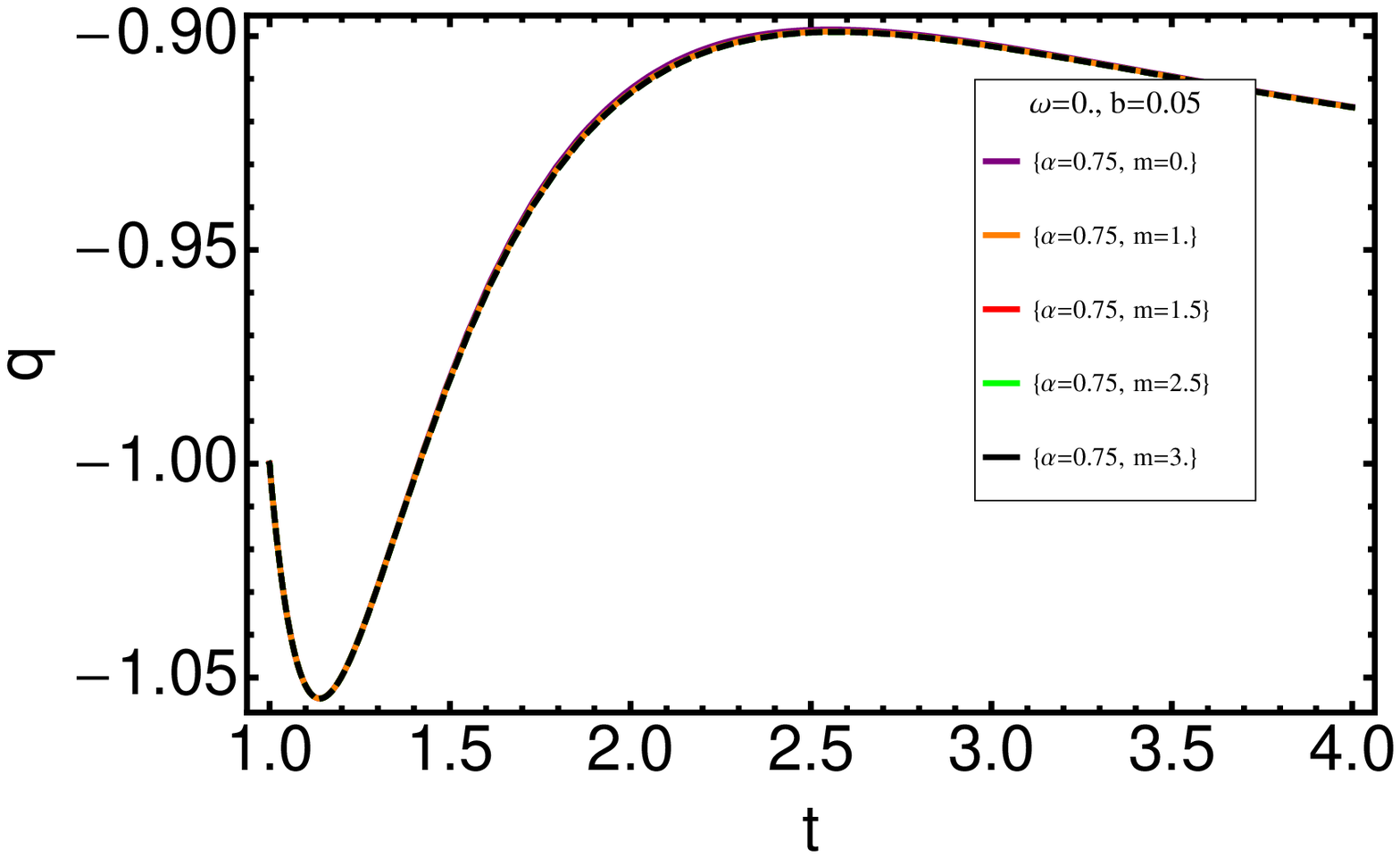}
 \end{array}$
 \end{center}
\caption{Behavior of Hubble parameter $H$ and $q$ against $t$ for ($t, H, \rho$)-dependent $\Lambda$ and the model 2.}
 \label{fig:7}
\end{figure}

\begin{figure}[h!]
 \begin{center}$
 \begin{array}{cccc}
\includegraphics[width=50 mm]{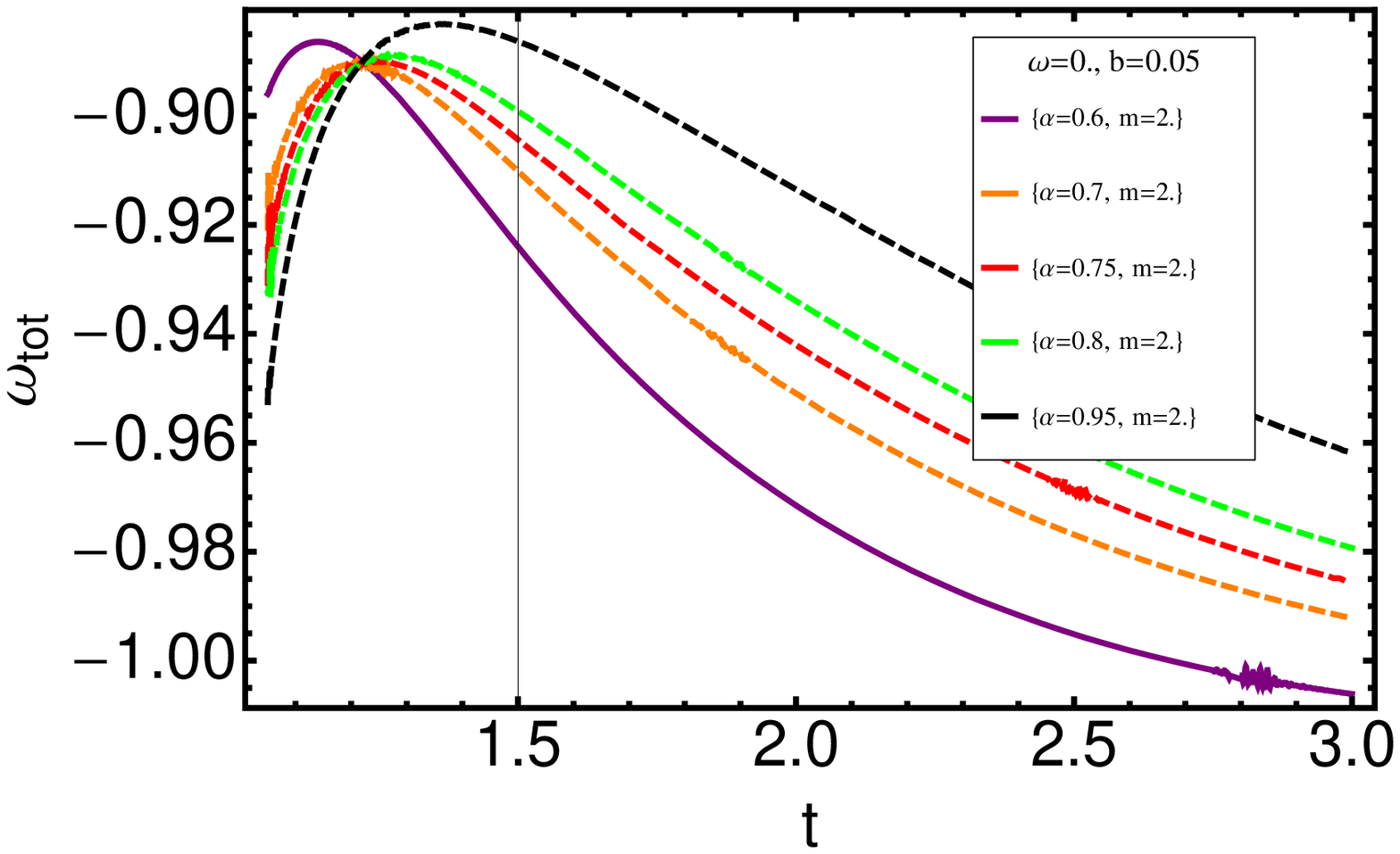} &
\includegraphics[width=50 mm]{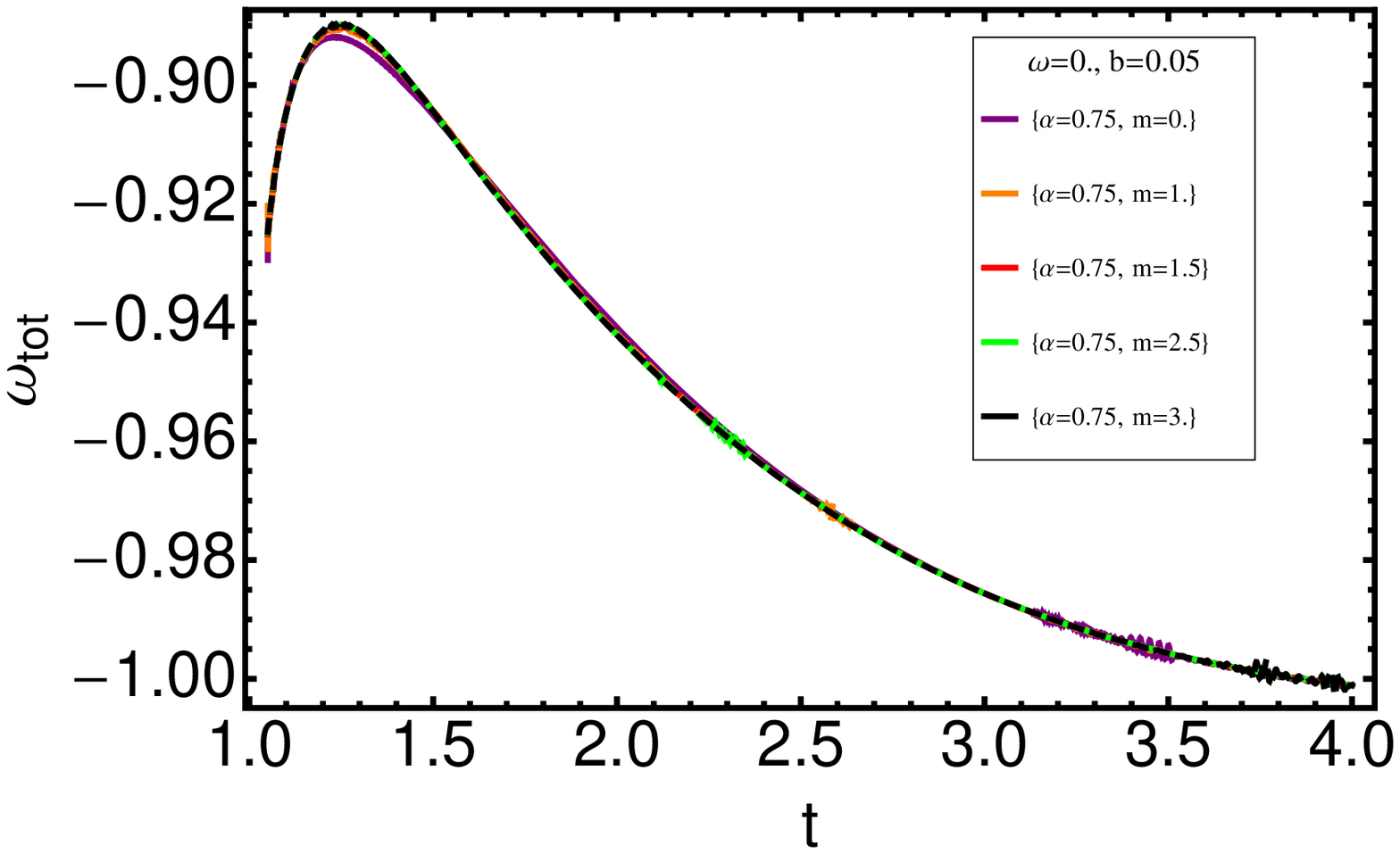}\\
\includegraphics[width=50 mm]{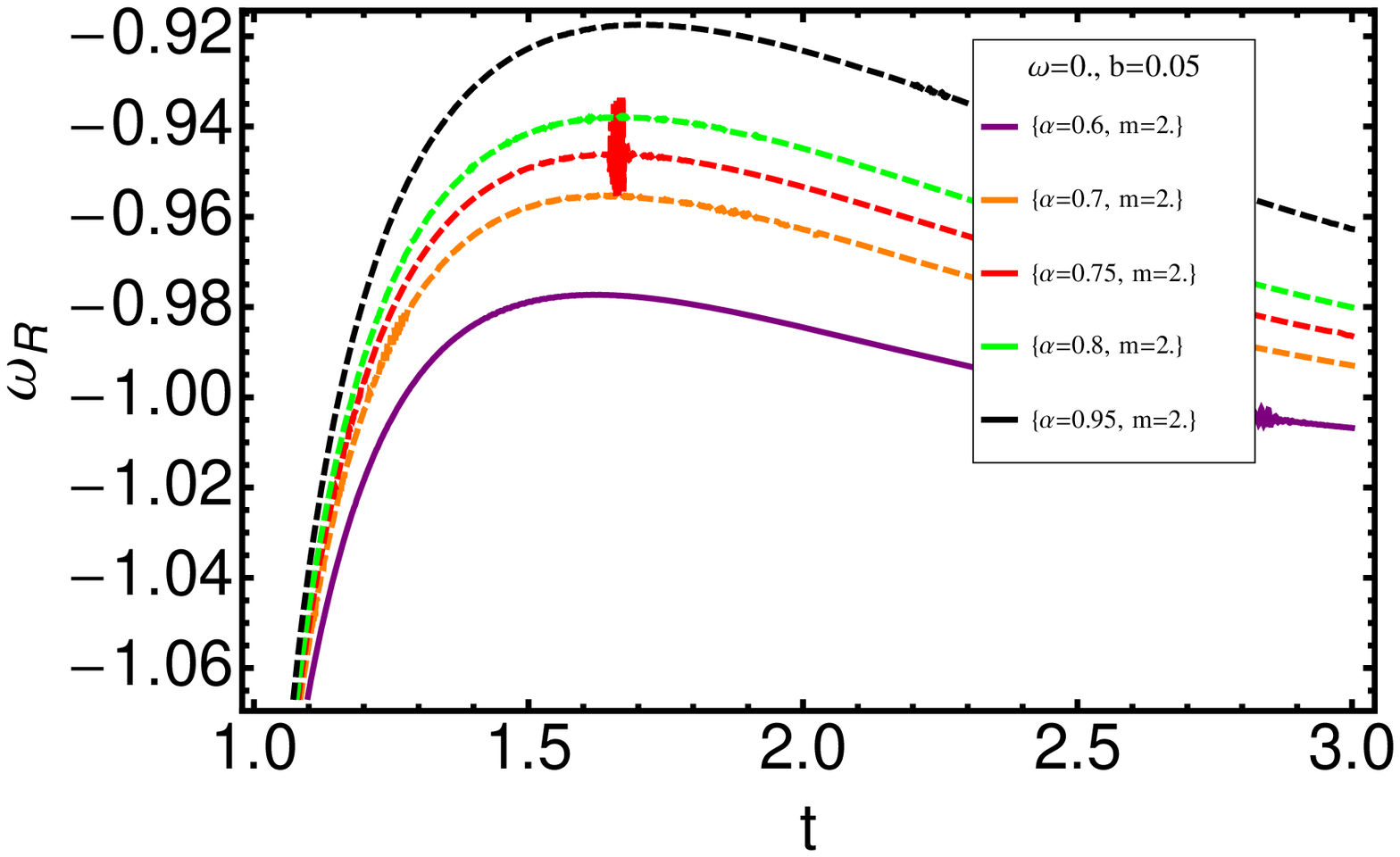}&
\includegraphics[width=50 mm]{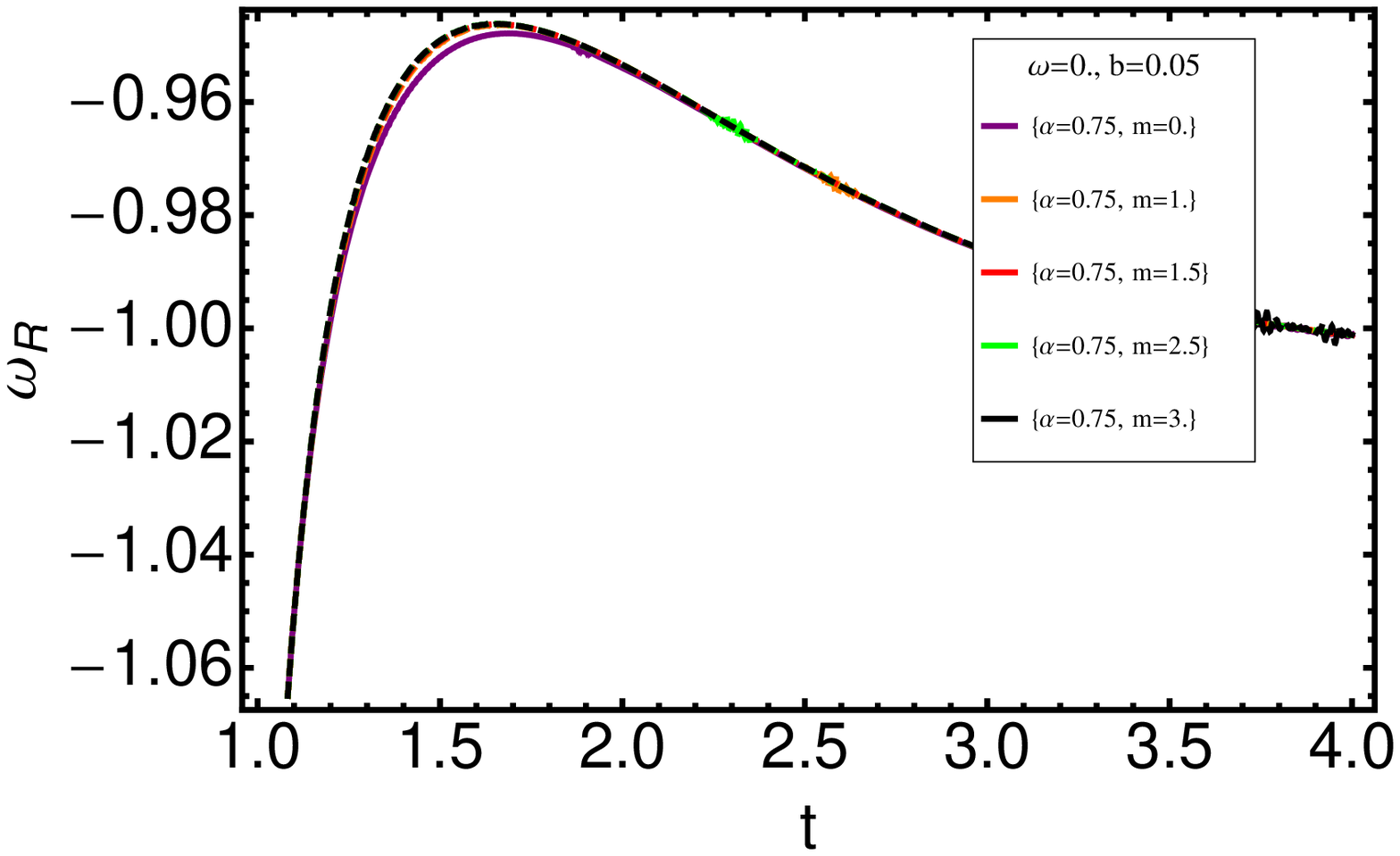}
\end{array}$
 \end{center}
\caption{Behavior of EoS parameter $\omega_{tot}$ and $\omega_{R}$ against $t$ for ($t, H, \rho$)-dependent $\Lambda$ and the model 2.}
 \label{fig:8}
\end{figure}

\section{Conclusion}
In this paper, we suggested a toy model of universe which filled with barotropic dark matter and Ricci
dark energy in Lyras geometry. Also we assumed the cosmological constant may vary with the density, Hubble parameter or explicitly by time. Interaction between dark matter and dark energy also considered in this paper. We considered 6 different models by choosing different kinds of interaction term and $\Lambda$. By using numerical analysis, we investigated important cosmological parameters of the models such as equation of state,
Hubble and deceleration parameters in terms of time. We compared our results with some observational data and concluded that the consideration of varying $\Lambda$ may be useful to obtain more agreement results with observations. Two first models based on constant $\Lambda$. Two second models based on $\Lambda=\rho$ which the first one (model 1) has agreement with BAO/CMB observations combining with SNIa data. Finally the last two models based on ($t, H, \rho$)-dependent $\Lambda$, and the first one (model 1) tells that the universe is in phantom phase from beginning to end, but the second one (model 2 ) indicates that the universe is in phantom phase initially and is in quintessence pase after late time.

\end{document}